\documentclass[useAMS,usenatbib,aas_macros]{mn2e}
\usepackage{graphicx}
\input{psfig.sty}
\usepackage{color}

\newcommand{\be}{\begin{equation}}
\newcommand{\ee}{\end{equation}}
 
\def\no{\noindent}

\newcommand{\msun}{M_\odot}

\newcommand{\ifm}[1]{\relax\ifmmode#1\else$\mathsurround=0pt #1$\fi}
\newcommand{\kms}{\ifmmode\,{\rm km}\,{\rm s}^{-1}\else km$\,$s$^{-1}$\fi}

\newcommand{\mpc}{\,{\rm Mpc}}

\newcommand{\ltsima}{$\; \buildrel < \over \sim \;$}
\newcommand{\lsim}{\lower.5ex\hbox{\ltsima}}
\newcommand{\gtsima}{$\; \buildrel > \over \sim \;$}
\newcommand{\gsim}{\lower.5ex\hbox{\gtsima}}

\def\M11{M_{11}}
\def\V100{V_{100}}
\def\R1{R_{Mpc}}
\def\T6{T_6}

\title[The star-forming progenitors of massive red galaxies] 
{The star-forming progenitors of massive red galaxies}

\author[A. Cattaneo, J. Woo, A. Dekel, S.M. Faber]
{{A.~Cattaneo$^{1\star}$, J.~Woo$^{2}$}, A.~Dekel$^{2}$, S.~M.~Faber$^3$\\
  $^1$Laboratoire d'Astrophysique de Marseille, UMR 6110 CNRS, Univ. d'Aix-Marseille, 38 rue F. Joliot-Curie, 13388 Marseille cedex 13, France \\
  $^2$Racah Institute of Physics, The Hebrew University, Jerusalem 91904, Israel \\
  $^3$UCO/Lick Observatory, University of California, USA\\
  $^\star${andrea.cattaneo@oamp.fr}\\}
\begin{document}

\pagerange{\pageref{firstpage}--\pageref{lastpage}} \pubyear{2012}

\maketitle

\label{firstpage}


\begin{abstract}
  The link between massive red galaxies in the local Universe and
  star-forming galaxies at high redshift is investigated with a
  semi-analytic model that has proven successful in many ways, e.g. 
  explaining the
  galaxy colour-magnitude bimodality and the stellar mass-age relation
  for red-sequence galaxies. 
 The model  is used to explore
 the processes that drive star formation in different types of galaxies as a function of stellar mass and redshift.  
 We find that most $z=2-4$ star-forming
  galaxies with $M_*>10^{10}M_\odot$ evolve into red-sequence galaxies. Also,
  most of the massive galaxies on the red-sequence today have passed through a phase of intense star formation at $z>2$.
  Specifically,  $\sim 90\%$ of today's red galaxies with $M_*>10^{11}M_\odot$
  were  fed during this phase
  by cold
  streams including minor mergers. Gas-rich major mergers are rare and the effects of merger-driven
  starbursts are ephemeral. On the other hand, major mergers are important in powering the most extreme starbursts.  Gas-rich mergers also explain the tail
  of intermediate-mass red galaxies that form relatively late, after
  the epoch of peak star formation.
  In two thirds of the currently red galaxies that had an intense star formation event at $z<1$, this event was triggered by
 a merger.
 
\end{abstract} 

\begin{keywords}
{
galaxies: ellipticals ---
galaxies: evolution ---
galaxies: formation ---
galaxies: high-redshift ---
galaxies: starbursts
}
\end{keywords}


\section{Introduction}

Sub-mm observations have uncovered a new population of high-$z$ star-forming galaxies (SFGs), 
which are detected through the far-infrared emission of dust heated by UV-bright newborn stars
\citep{smail_etal97,eales_etal99}.  {These lie along a sequence in
  the star formation rate (SFR)-stellar mass plane that is elevated with respect to the
  $z=0$ relation \citep{noeske_etal07b, elbaz_etal11}. 

In the local Universe, ultraluminous infrared galaxies (ULIRGs){,
  whose SFRs are several times higher than local SFGs of the same
  mass,} are directly linked to mergers
\citep{sanders_etal86}. Hydrodynamic simulations reproduce this
observational finding \citep{mihos_hernquist94,barnes_hernquist96,mihos_hernquist96}. The same simulations
also find that the morphological properties of the remnant of gas-rich
mergers are consistent with those of $L_*$ ellipticals
\citep{barnes_hernquist91,cox_etal06}. These observational and
theoretical results have been used to support the merger scenario, in
which elliptical galaxies are formed by mergers of spirals
\citep{toomre_toomre72,hopkins_etal06}. In this scenario, ULIRGs
correspond to the formation of elliptical galaxies caught in the act.

Massive red galaxies (i.e., Es and S0s) formed the bulk of their stars
at high $z$ \citep{thomas_etal05}, where conditions were different
from the local Universe.  Therefore, it is not straightforward to
generalise the lesson from local ULIRGs to higher redshifts. However, progress in astronomical
instrumentation has made it possible to observe the morphologies and
kinematics of {the high-$z$ counterparts of ULIRGs. They are galaxies that are
detected in the
  sub-mm and that
 form stars at rates of hundreds of solar masses per
  year.  We refer to high-$z$ ULIRGs as  sub-mm galaxies (SMGs),
  independently of their actual sub-mm fluxes.

\citet{engel_etal10} have used CO interferometric data to conclude
that most bright ($L_{\rm IR}>5\times 10^{12}L_\odot$) SMGs are major
mergers.  \citet{foersterschreiber_etal09} have observed a larger
sample of less extreme objects with integral field spectroscopy. They
have found that only one third of their galaxies are interacting or merging. Another one third
are rotation-dominated turbulent discs and the rest are
velocity-dispersion-dominated objects.  \citet{kartalpepe_etal12}
present further evidence that the highest-SFR galaxies tend to be
highly disturbed even at $z\sim 2$.

These observations have a natural explanation in a scenario
in which cold streams in massive haloes are the main mode of galaxy
mass accretion \citep{keres_etal05,dekel_birnboim06}.  {The cosmological simulations and theoretical analysis
  of \citet{dekel_etal09}} show that only one third of the stream mass
is in clumps leading to mergers of mass ratios greater than 1:10. The
rest is in smoother flows that preserve the disc's rotation.  However,
discs formed by fast accretion of cold streams are turbulent and
violently unstable.  They break into clumps that coalesce into a
central spheroid, hence providing an alternative to the merger
scenario for the formation of spheroids \citep{elmegreen_etal08}.
\citet{dekel_etal09} find that most ($\sim 3/4$) high-$z$ SFGs are
stream-fed and that mergers are only necessary to explain the
brightest SMGs, the latter being the highest-SFR subclass within SFGs.

The question is which of these two types of objects are the
progenitors of $z=0$ massive red galaxies, and conversely, 
 {how many} $z=0$ massive red galaxies have such progenitors.
  We should also like to know
  what proportion of high-$z$ SFG's ultimately turn into
  massive red galaxies.  In this
article, we address these questions with a semi-analytic model that has proved
successful in explaining local data, such as the galaxy colour - magnitude distribution in
the Sloan Digital Sky Survey \citep{cattaneo_etal06} and the stellar
mass - age relation for red-sequence galaxies \citep{cattaneo_etal08}.
The key feature that makes the model successful is 
inclusion of multiple mechanisms (explained below) whereby star formation can be quenched in SFGs, 
causing them to move to the red sequence.  
The model thus provides a plausible laboratory in which the importance of various gas accretion and 
quenching mechanisms for different types of galaxies can be compared with one another.
 
 In this paper,
the model is used both to evolve the properties of high-$z$ SFGs to low $z$
and to trace the progenitors of today's massive red galaxies back in time.  It
is necessary to analyze the problem both ways because even if all high
redshift galaxies of one type evolved into massive red galaxies, that
would not imply that all massive red galaxies derived from that type
of object. We shall also see how the model can be used to separate
stream-fed and merger-driven star formation.

It is straightforward to evolve galaxies from a given $z$ to $z=0$,
once a semi-analytic model is given.  It is trickier to tell
what type of object are the progenitors of the local population of massive red galaxies
because the properties of the progenitors of massive red
galaxies are time-dependent.  The main progenitor of a $z=0$
elliptical may be a clumpy disc at $z=4$, a major gas-rich merger at
$z=3$, and a massive red object at $z=2$.

We resolve this ambiguity by introducing the redshift $z_{\rm peak}$
at which the star formation rate (SFR) of a galaxy's main progenitor
has an absolute maximum.  Stream-fed star formation is much more
continuous than merger-driven star formation. If it dominates at
$z_{\rm peak}$, it dominates at all times.  In this case, star
formation is unambiguously stream-fed.  The problem is more
complicated when the peak SFR is linked to a merger event.  In that
case, we need to investigate the event's contribution to the build-up
of the final stellar mass.

The plan of this work is thus as follows. In Section~2, we recall the
main assumptions of the GalICS semi-analytic model used for this work.
Our goal is not to give a detailed description of the GalICS model (it
can be found in \citealp{hatton_etal03} and
  \citealp{cattaneo_etal06}).  It is rather to make clear how we separate
stream-fed and merger-driven star formation.  
We shall see that, in our model, stream-fed gas accretion and minor merger contribute to both disc star formation
(directly) and bulge star formation (indirectly, by triggering disc instabilities), while major mergers only contribute
the triggering of bulge star formation.
Therefore, the relative importance of bulge star formation can be used to put an upper limit to the importance of major
mergers in the star formation histories of galaxies.
In Section~3, 
we describe how we assign a galaxy descendent population to a parent one, and vice versa.
In Section~4, we present
the mass -- SFR relation at $z=2,\,3,\,4$, and we discuss
which subset of the galaxies that populate this relation ends up on
the red sequence at $z=0$.  In Section~5, we discuss the progenitors of massive red galaxies,
show some
characteristic SFR histories, and describe how the properties at
$z_{\rm peak}$ ($z_{\rm peak}$, SFR$_{\rm peak}$, driving mechanism)
depend on a galaxy's stellar mass at $z=0$.  Section~6 presents the
conclusions of the article.


\section{The galaxy formation model} 
\label{sec:galics} 

GalICS (Galaxies In Cosmological Simulations; \citealp{hatton_etal03})
is a method to simulate the formation of galaxies in a $\Lambda$CDM
Universe.  It combines cosmological N-body simulations of the
gravitational clustering of the dark matter with a semi-analytic (SAM)
approach to the physics of the baryons (gas accretion, galaxy mergers,
star formation and feedback).

The version of GalICS used here is the same as the `new model'
introduced in \citet{cattaneo_etal06} and \citet{cattaneo_etal08} and
uses the same parameter values.  We now recall its fundamental
assumptions. We concentrate on the points that are relevant for this
work and refer to the articles above for a more detailed description.

\subsection{Dark-matter simulation}

The cosmological N-body simulation that follows the hierarchical
clustering of the dark-matter component was carried out with a
parallel tree code. It assumes a flat $\Lambda$CDM Universe with a
cosmological constant of $\Omega_{\Lambda}=0.667$ and a Hubble
constant of $H_0=66.7{\rm\,km\,s}^{-1}$.  The $\Lambda$CDM power
spectrum of initial fluctuations is normalized to $\sigma_8 =0.88$.
The computational volume is a cube of {size} $(150{\rm\,Mpc})^3$ with
$256^3$ particles of $8.3\times 10^9$M$_{\odot}$ each and a smoothing
length of $29.3\,$kpc.  The simulation produced 100 output snapshots
spaced logarithmically in expansion factor $(1+z)^{-1}$ from $z=35.59$
to $z=0$.

Each snapshot was analysed with a friends-of-friends algorithm
\citep{davis_etal85} to identify virialized haloes containing more
than 20 particles.  The minimum halo mass is thus $1.65\times
10^{11}M_\odot$.

Merger trees are constructed by linking the haloes identified in each
snapshot with their progenitors in the previous snapshot, i.e., all
predecessors from which the halo has inherited one or more particles.

\subsection{Disc formation at the centre of dark-matter haloes}

The baryons in each halo are assumed to cool efficiently, i.e. to
stream cold, onto the central galaxy until the dark matter halo
reaches a critical mass of
\begin{equation}
M_{\rm crit}=M_{\rm shock}\times {\rm max}\{1,\,10^{1.3(z-z_{\rm c})}\}.
\label{eq:mcrit} 
\end{equation}
This assumption has a simple physical justification.  Above a critical
halo mass of $M_{\rm shock}\sim 10^{12}M_\odot$, the gas that streams
into the halo is shock heated \citep{dekel_birnboim06} and becomes
vulnerable to black-hole heating, which prevents it from cooling down
again (see \citealp{cattaneo_etal09} for a review; also see
\citealp{dekel_birnboim08} and \citealp{khochfar_ostriker08} for a
discussion of the contribution of gravitational heating).

The effects of changing $M_{\rm crit}$ have been studied quantitatively in \citet{cattaneo_etal06}, where we
have shown that the value of this parameter is constrained within a factor of two.

The term $\propto 10^{1.3z}$ was introduced to account for the more efficient
penetration of cold streams in massive haloes at high redshift (see
\citealp{dekel_birnboim06} and \citealp{dekel_etal09}).

In \citet{cattaneo_etal06}, we set the model parameters to $M_{\rm
  shock}= 2\times 10^{12}M_\odot$ by fitting the colour-magnitude
distribution in the Sloan Digital Sky Survey and to $z_{\rm c}=3.2$ by
fitting the Lyman-break galaxy luminosity function at $z\simeq 3$.  
The gas that falls to the centre settles into a disc.  Star formation
is activated when the gas surface density is $\Sigma_{\rm
  gas}>10^{20}{\rm\,H}{\rm\,cm}^{-2}\simeq 1\,M_\odot{\rm\,pc}^{-2}$. The star formation rate is described by the star formation law:
\begin{equation}
\label{eq:sfr} 
\dot{M}_{\rm star}={M_{\rm cold}\over\beta_*t_{\rm dyn}}(1+z)^{\alpha_*},
\end{equation}
where $M_{\rm cold}$ is the mass of cold star-forming gas and $t_{\rm
  dyn}$ is the half-rotation time at the disc's half-mass radius
(determined assuming angular momentum conservation and an exponential
profile). The free parameters $\beta = 50$ and $\alpha=0.6$ were fixed
by fitting the Kennicutt law (see \citealp{guiderdoni_etal98}) and the
luminosity function of Lyman-break galaxies, respectively. 
It is important to be aware that our predictions for the luminosity function
of Lyman-break galaxies at $z\simeq 3$ are highly sensitive to our dust extinction calculations, which contain large uncertainties, and
that these uncertainties trickle down in the best-fit values for both $\alpha$ and 
$z_{\rm c}$.

In a recent article,  \citet{krumholz_etal12} studied how Eq.~(\ref{eq:sfr}),
which is equivalent to the Kennicutt law for local discs,
extrapolates to high-redshift discs and starburst.
They found that Eq.~(\ref{eq:sfr}) applies to all galaxy types and at all redshifts with $\alpha_*\sim 0$ if $t_{\rm dyn}$ is the {\it local} freefall time. 
However, if $t_{\rm dyn}$ is the {\it global} dynamical time, i.e. the half-rotation time, then
Eq.~(\ref{eq:sfr}) extends to high-redshift discs, but underestimates the SFR for both low- and high-redshift starbursts.
Therefore, the local freefall time is a better estimator of the star formation timescale than the global dynamical time.
However, in our model we are obliged to use the global time because 
 our model does not consider the vertical structure of galactic discs, which is important for the local freefall time.

As Eq.~(\ref{eq:sfr}) with $\alpha_*\sim 0$ extends reasonably well to high-$z$ galaxies classified as discs, one may argue that there is no ground for allowing $\alpha_*>0$. However, if many of the high-$z$ galaxies classified as starbursts are starbursting discs (very much like the local galaxy M82), then there is indeed a case to do so.
In that case, boosting the mean star-formation efficiency of discs at high $z$ would be a way to account 
for the fact that, at high $z$, a much greater fraction of the disc population is in a starbursting rather than a quiescent mode.
Notice, however, that, for $\alpha = 0.6$, the term $(1+z)^\alpha$ boosts the SFRs at $z\sim 3$ by little more than a factor of two, and that this is
less than the precision with which the normalisation of the Kennicutt law can be determined at high $z$.
In the Discussion, we shall consider how the uncertainty on $z_{\rm c}$ and $\alpha_*$ may affect the conclusions of the article.

As smooth gas accretion is the only mechanism through which discs are
allowed to acquire gas, disc star formation is always stream-fed star
formation.
This point will be of great importance for the interpretation of our results.

\subsection{Bulges}

In GalICS, bulges are formed by two mechanisms: mergers and violent
disc intabilities.
  
Bulge growth via disc instabilities is modelled by assuming that the
bulge mass increases until it is sufficiently large to stabilize the
disc.  The stability criterion is
\begin{equation}
\sqrt{ {0.5{\rm G}M_{\rm disc}\over r_{1/2}}  }< \eta \sqrt{  {{\rm G}M(r_{1/2})\over r_{1/2}} },
\end{equation}
where $M_{\rm disc}$ is the disc mass, $r_{1/2}$ is the exponential disc's half-mass radius, 
$M(r_{1/2})$ is the total mass within the disc half-mass radius, and $\eta$ is a parameter that controls the
instability threshold (see e.g. \citealp{vandenbosch98}). This criterion translates into the condition
\begin{equation}
\sqrt{0.5M_{\rm disc}}<
\eta \sqrt{0.5M_{\rm disc}+M_{\rm bulge}+M_{\rm dm}(r_{1/2})},
\label{discinst}
\end{equation}
where $M_{\rm dm}(r_{1/2})$ is the mass of the dark matter within the 
disc's half-mass radius.
The instability threshold is set to $\eta=0.7$, as in \citet{hatton_etal03}.
For this value of $\eta$,
blue galaxies have bulges, yet they remain largely disc-dominated
objects.
The influence of the $\eta$ parameter on our results will be discussed in Section~6 (Discussion and Conclusion).

Eq.~(\ref{discinst}) implies that, in the absence of mergers, the
growth of the bulge is directly linked to the growth of the disc.
Since bulges formed by violent disc instabilities never dominate the
total galaxy mass in our model, bulge star formation triggered by disc
instabilities is always secondary with respect to disc star
formation. This is true irrespective of the star formation law. Even
if bulge star formation happens on a much shorter timescale than disc
star formation (see below), gas cannot be converted into stars faster
than it is supplied to the bulge.  In a context where the bulge gas
accretion rate is more or less proportional but always lower than the
disc gas accretion rate, disc star formation is bound to be dominant.

In GalICS, galaxy mergers are mainly due to orbital decay of satellite
galaxies through dynamical friction after halo mergers have resulted
in the formation of galaxy groups and clusters.  The fraction of the
disc that is transferred to the bulge in a merger grows with the mass
ratio of the merging galaxies. It ranges from zero for a very minor
merger to unity for an equal mass merger, which means that bulge
formation is essentially linked to major mergers.  The bulge that
forms is assumed to have a \citet{Hernquist90} profile and its radius
is determined based on an energy conservation argument.  This
simplified picture of the dynamics of morphological transformations
is, nevertheless, consistent with key observational constraints such
as the Faber-Jackson relation and the Fundamental Plane of spheroids
\citep{hatton_etal03}.

Readers who are familiar with the GalICS model will know that, in the
code, gas is not moved directly from the disc to the bulge but it
passes through a transitional component called the starburst.  The star
formation law has exactly the same form (Eq.~\ref{eq:sfr}) for all
three components, but the starburst is assumed to have scale length
that is equal to one tenth of the disc scale-length.  Therefore, the
starburst dynamical time that enters Eq.~(\ref{eq:sfr}) is ten times
shorter than the bulge dynamical time, which implies a factor of ten
increase in the SFR.  Stars formed in the starburst are moved to the
bulge after they have reached an age of $100\,$Myr and the only gas in
the bulge is that from stellar mass loss.  However, it is not
necessary to enter this level of detail to understand any of the
conclusions of this article. We shall therefore simply speak of discs
and bulges, where by bulge we mean the sum of the bulge and the
starburst.

It follows from our discussion that while disc star formation is
always stream fed, bulge star formation can be caused by two
processes, since both gas-rich mergers and violent disc instabilities (such as
those deriving from fast accretion of cold streams)
contribute to bulge star formation.  This point is
important for the interpretation of our results and will be discussed
further later on.

\subsection{Feedback}

We conclude our presentation of the GalICS model with a brief
description of how it handles stellar evolution and feedback.

Stars are evolved between snapshots using substeps of at most 1\,Myr.
During each sub-step, stars release mass and energy into the
interstellar medium. Most of the mass comes from the red giant and the
asymptotic giant branches of stellar evolution, while most of the
energy comes from shocks due to supernova explosions. The enriched
material released in the late stages of stellar evolution is mixed
with the cold phase, while the energy released from supernovae is used
to reheat the cold gas and to return it to the hot phase in the halo
(Eq.~\ref{eq:mdot}).  Reheated gas is ejected from the halo if the
potential is shallow enough.  The rate of mass loss through
supernova-driven winds $\dot{M}_{\rm w}$ is determined by the equation
\begin{equation} 
{1\over2}\dot{M}_{\rm w}v_{\rm esc}^2
=\epsilon_{\rm SN}\eta_{\rm SN}E_{\rm SN}\dot{M}_{\rm star},
\label{eq:mdot}
\end{equation}
where $E_{\rm SN}=10^{51}\,$erg is the energy of a supernova,
$\eta_{\rm SN}=0.0093$ is the number of supernovae for $1\,M_\odot$ of
stars formed and $v_{\rm esc}$ is the escape velocity
\citep{dekel_silk86}.

In GalICS, feedback is computed separately for each galaxy component.
We use $v_{\rm esc}\simeq 1.84v_{\rm c}$ for discs and $v_{\rm
  esc}=2\sigma$ for bulges/starbursts.  The supernova efficiency
$\epsilon_{\rm SN}\simeq 0.2$ is similar to that commonly adopted in
SAMs \citep{somerville_primack99,cole_etal00}.

\begin{figure*}
\noindent
\includegraphics[width=0.33\hsize]{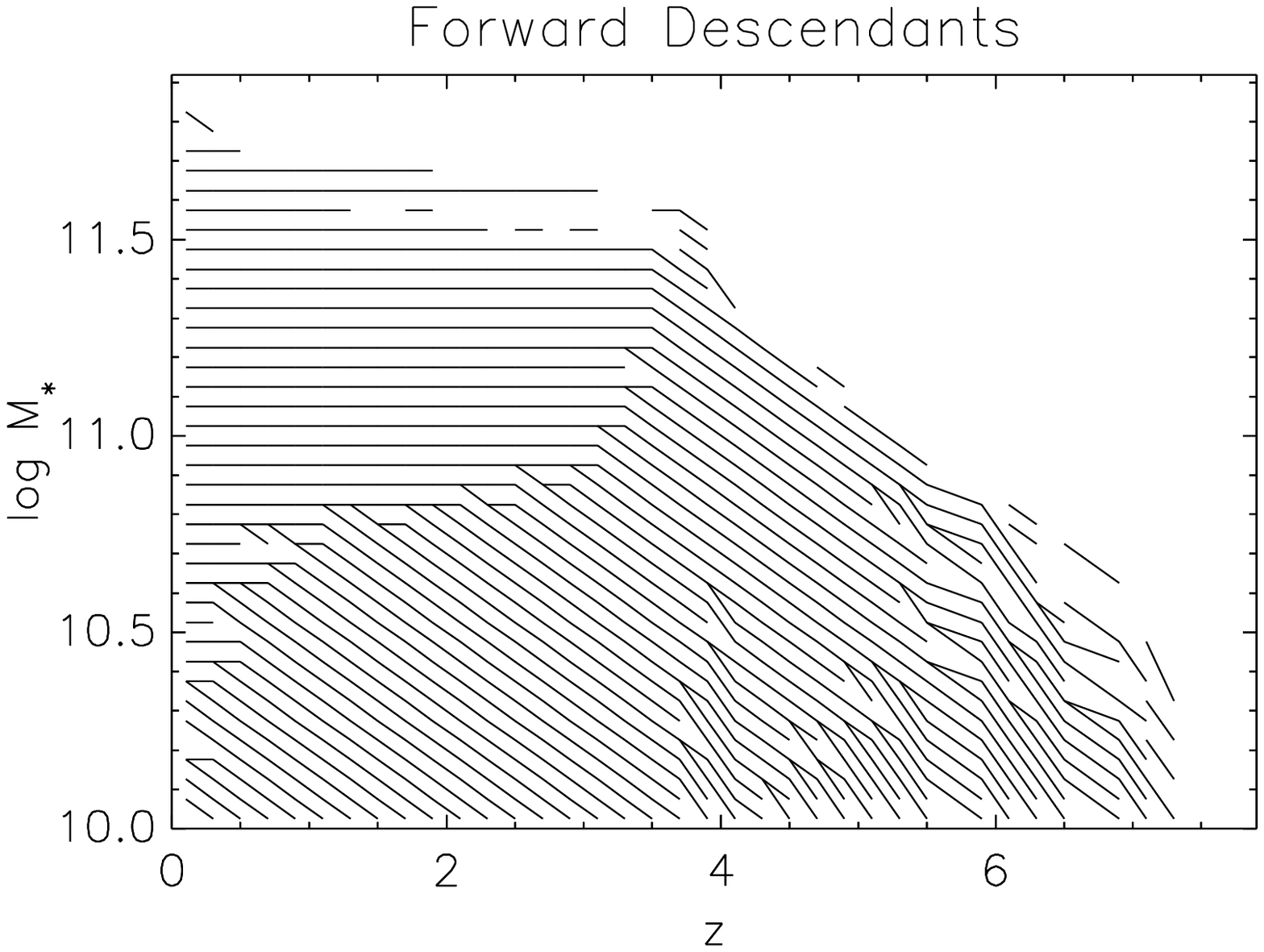}
\includegraphics[width=0.33\hsize]{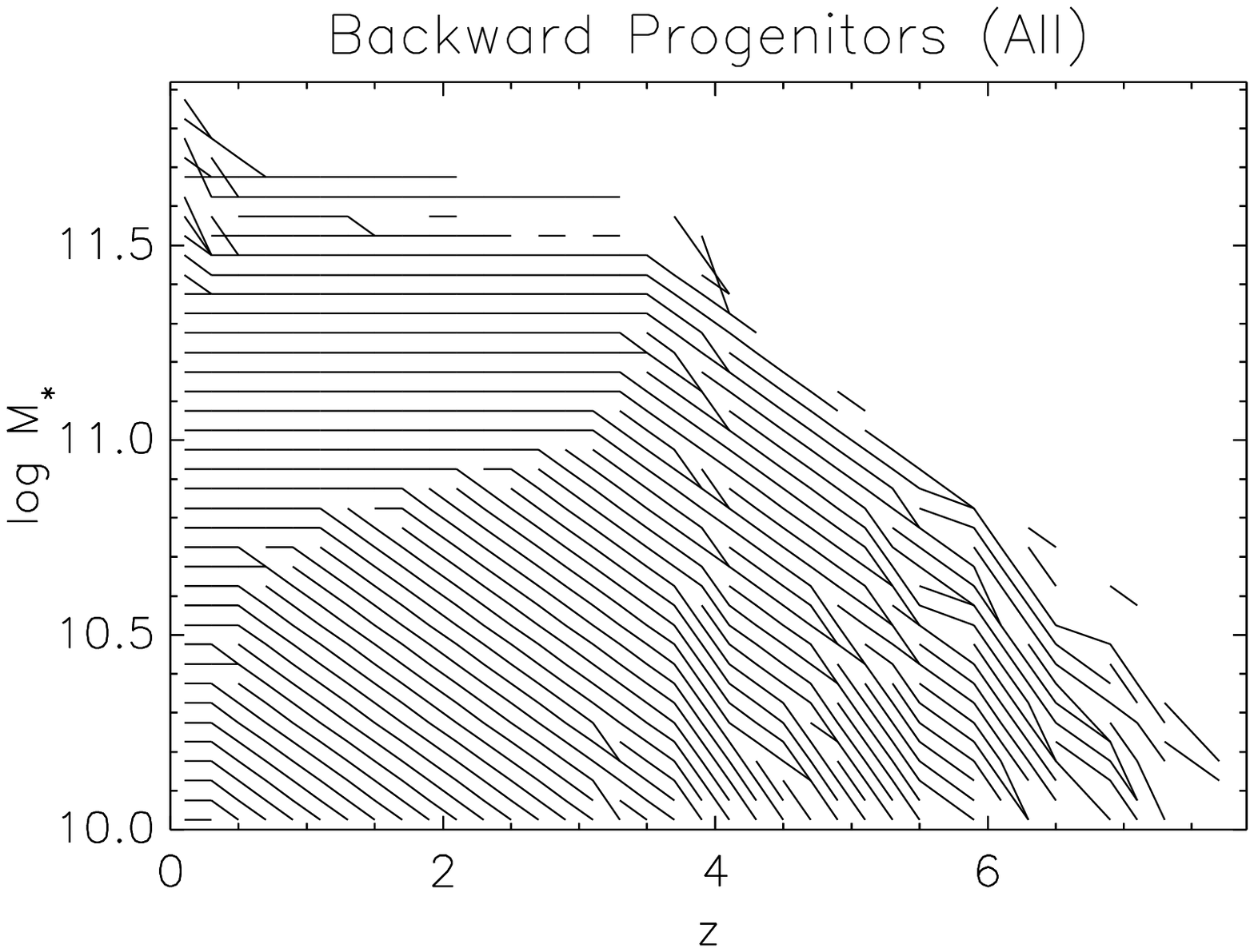}
\includegraphics[width=0.33\hsize]{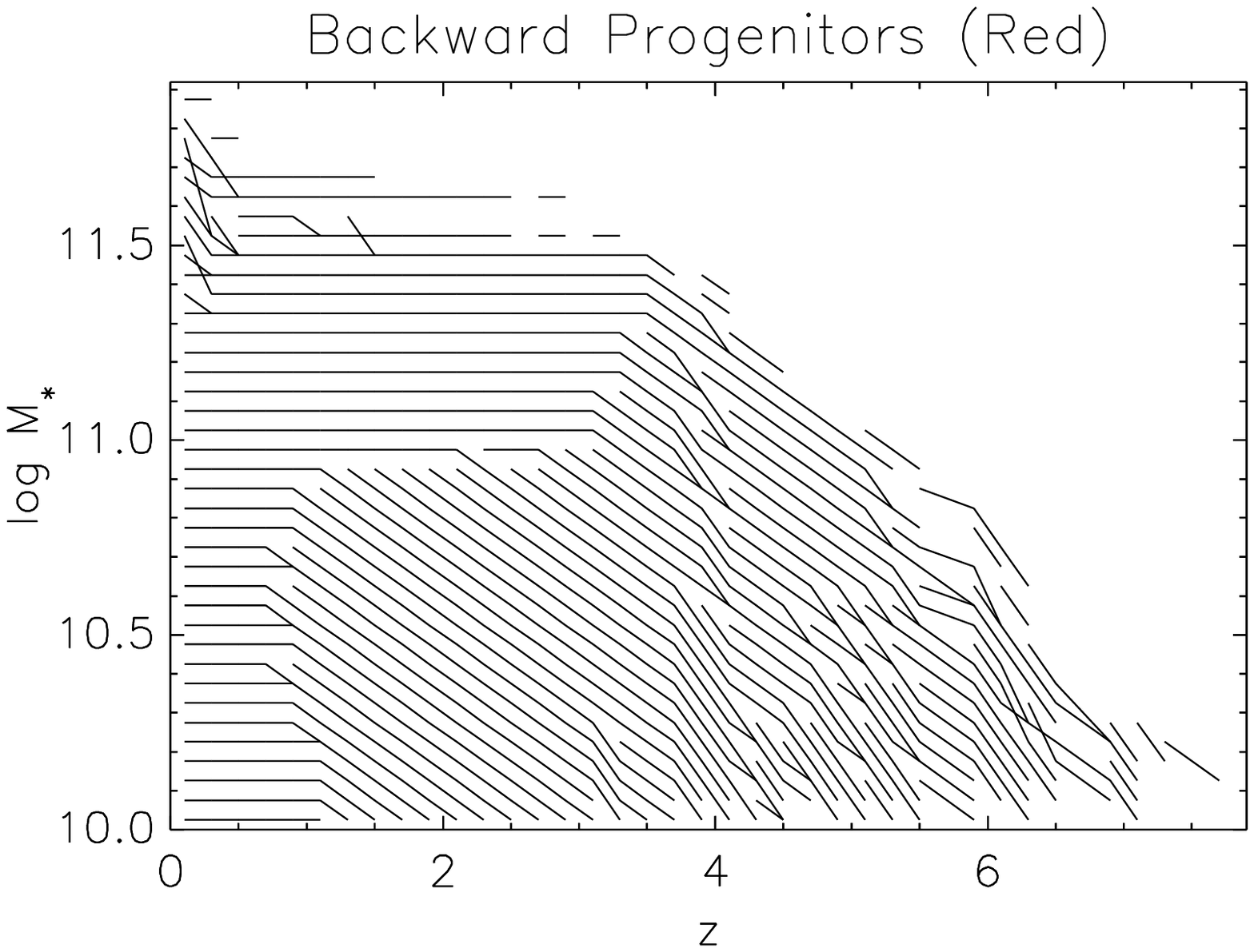}
\caption{Forward (left panel) and backward (central and right panels) evolutionary tracks of galaxies on the $M_*$ - $z$ plane.
The left panel and the central panel are for the entire population. The right panel is only for galaxies that are on the red sequence at $z=0$.
Forward evolutionary tracks are constructed by linking a tile with the tile corresponding to the median mass of the descendant galaxies.
Backward evolutionary tracks are constructed by linking a tile with the tile corresponding to the median mass of the main progenitors.}
\label{fig1new}
\end{figure*}

\subsection{Quenching of star formation}

As in Cattaneo et al. (2008),  in haloes above $M_{\rm crit}$, we do not just shut down gas accretion;
we also shut down star formation.
Moreover, we suppress gas accretion in galaxies
where the bulge mass is larger than half the total stellar mass.
As long as $M_{\rm halo}<M_{\rm crit}$, gas accretion can start again if this condition is no longer verified.

These further assumptions were introduced not based on a compelling physical argument,
(though some physical justifications are possible; see below),
but rather because we found that they improved the
agreement with the observed galaxy colour - magnitude distribution,
even though an acceptable fit could be obtained without them (Fig.~9 of Cattaneo
et al. 2006 shows how these further assumptions improve the basic model in which we simply shut down gas accretion
at $M_{\rm halo}>M_{\rm crit}$).

The second assumption implies that many ($>57\%$) red-sequence
galaxies are quenched following a merger event that has caused the
bulge mass to increase above the disc mass (note that the bulge mass never exceeds the disc mass by violent instability alone; 
see previous discussion).  This is particularly true
for galaxies around $M_*\sim 10^{11}\,M_\odot$ (Cattaneo et al. 2008,
Fig.~3).  

Two possible physical explanations for this behaviour are i) quenching induced by
quasar feedback following black hole growth activated by merging
(\citealp{springel_etal05b}; \citealp{hopkins_etal06}; also see
\citealp{cattaneo_etal09} for a review), and ii) morphological quenching as
proposed by \citet{martig_etal09}.  These authors presented a picture
in which a large bulge stabilizes disc instabilities and therefore
inhibits the formation of spiral arms, which are the main sites of
star formation in a spiral galaxy.  

In the first case, star formation is quenched because the cold gas is blown out.
In the second, the gas is not blown out, but it is prevented from making stars.
One should notice, however,
that shutting down gas accretion and shutting down star formation produce very similar effects
when plugged into our semianalytic model.

\section{Bridging past and present}

\begin{figure*}
\noindent
\includegraphics[width=0.33\hsize]{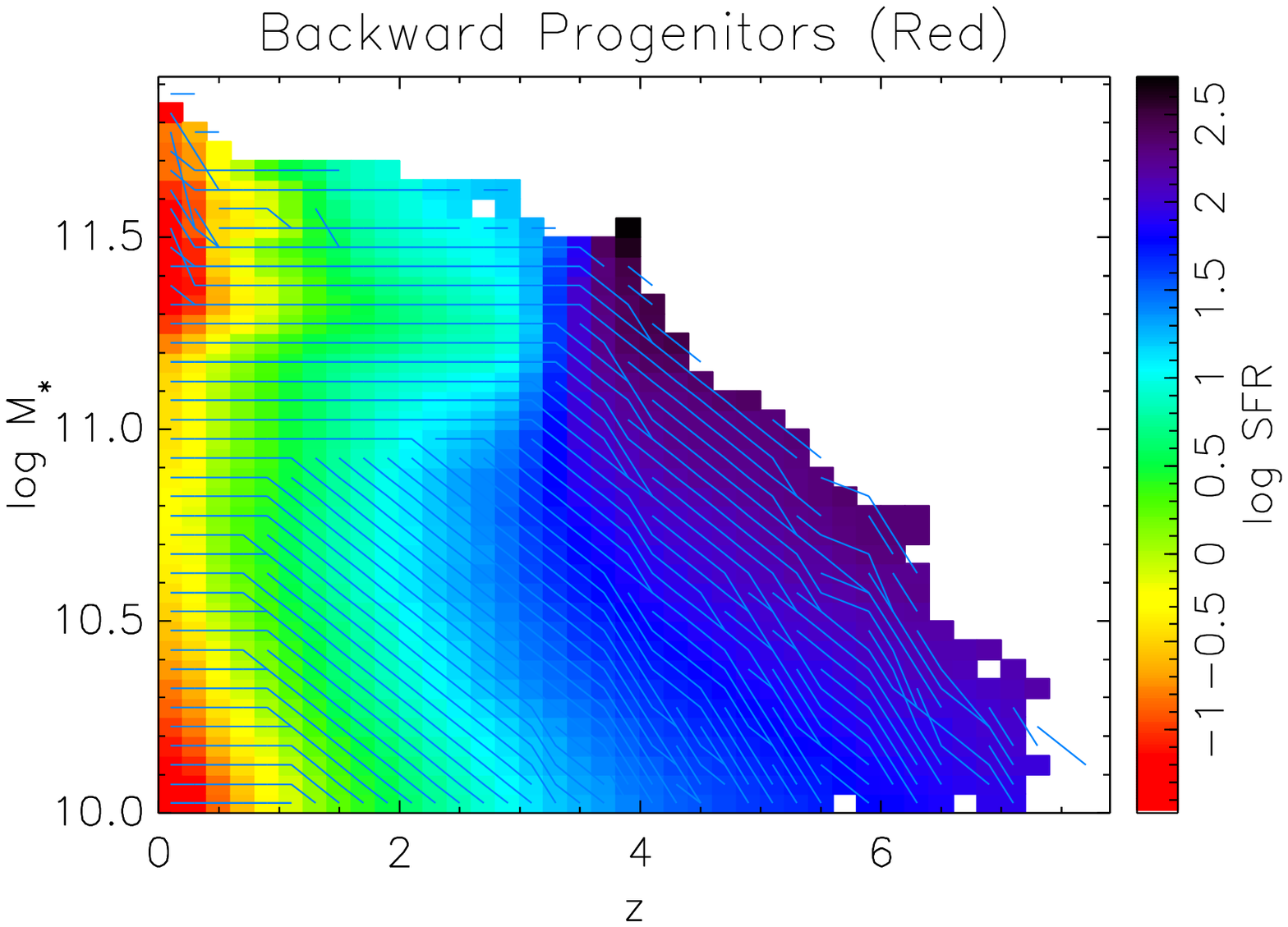}
\includegraphics[width=0.33\hsize]{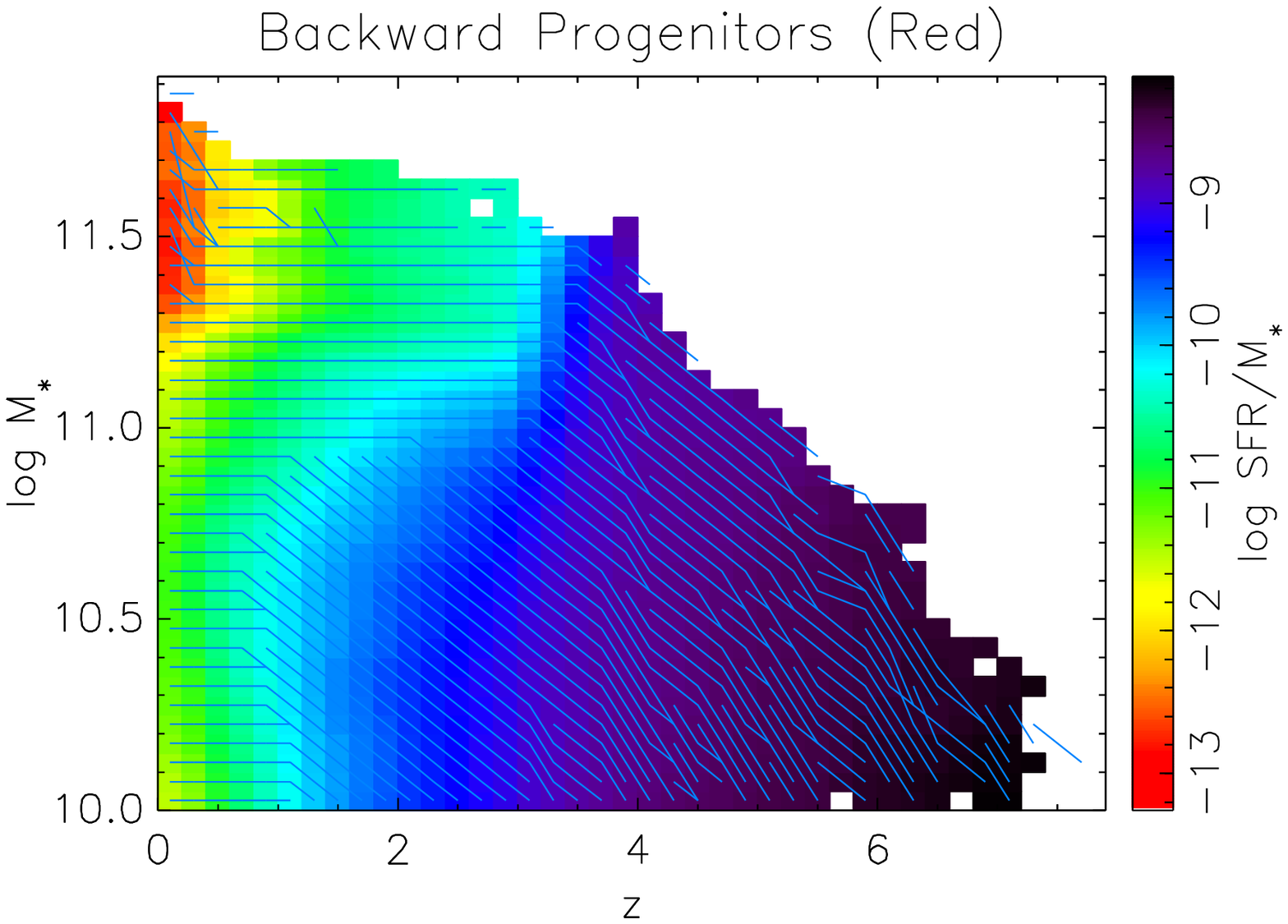}
\includegraphics[width=0.33\hsize]{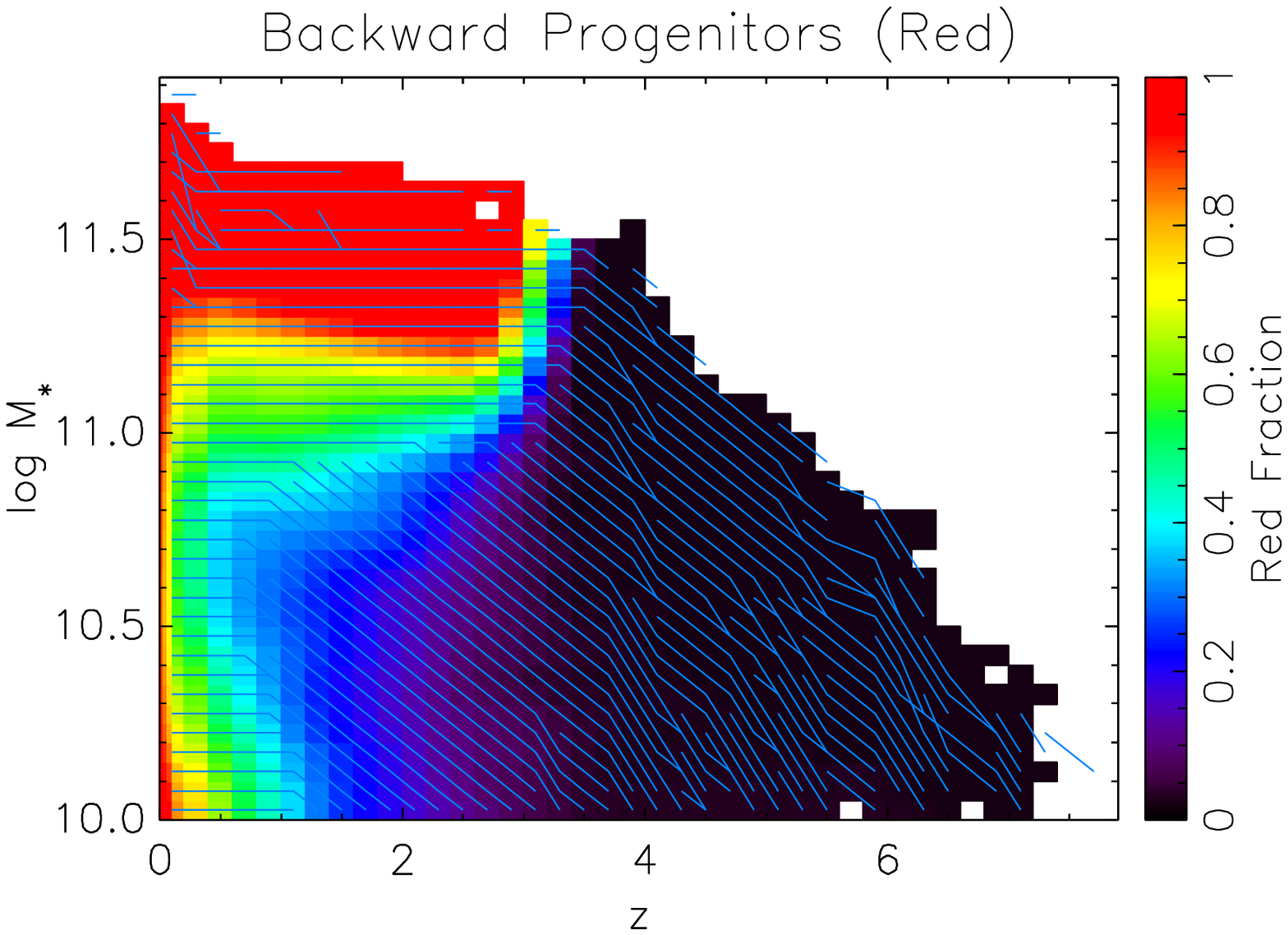}
\caption{Median SFR (left panel), specific SFR (centre panel) and fraction of red galaxies (right panel) on each tile of the $M_*$ - $z$ diagram.
The curves are the same backward evolutionary tracks as in the right panel of Fig.~\ref{fig1new}.}
\label{fig1bisnew}
\end{figure*}

Having described the model, we are now ready to explain how we analyse the results. As this work
is focussed on the nature of high-$z$ SFGs and the progenitors of low-$z$ red (early-type) galaxies, we need
a procedure to assign $z=0$ descendants to high-$z$ galaxies and high-$z$ progenitors to $z=0$ galaxies.
One way to do it is to follow the galaxy flow on a stellar mass - redshift diagram.

Fig.~\ref{fig1new} has been constructed by separating the stellar mass - redshift diagram into tiles.
For each tile, that is, for each a stellar mass and redshift interval,
we take the galaxies one by one and ask what is the median mass of their immediate descendants.
Once we have this information, we can use a trait to connect each tile to the tile in the following timestep 
that corresponds to the median descendant mass.
This is what we have in the left panel of Fig.~\ref{fig1new}, which shows the `forward' flow of galaxies on the $M_*$ - $z$ diagram.

The middle panel  is very similar, but it shows the `backward' flow.
Instead of looking for the median mass of the descendants of the galaxies on a tile, we look for the median mass of their main progenitors.
So we connect each tile with another tile at the previous timestep.

The right panel is identical to the middle panel, except  that it only traces back in time the progenitors of those galaxies that are red at $z=0$.

Let us start by comparing the left and middle panels. We
see that the forward and the backward evolution are not identical.
The reason for this difference is easily explained. It is possible that
most of the galaxies in two mass bins{, $M_1$ and $M_2$,} at
redshift $z_1$ come from {the} same mass bin{, $M_3$,} at the
previous redshift step $z_2>z_1$.  That means that some of the
galaxies in {$M_3$} at $z_2$ end up in {$M_1$} at $z_1$ and
some in {$M_2$}.  Let us assume that most end up in {$M_1$}.
{Then only $M_1$} at $z_1$ will be linked to {$M_3$} at $z_2$
in the forward evolution but both {$M_1$ and $M_2$} at $z_1$ will
be linked to {$M_3$} at $z_2$ in the backward evolution.

This is what happens, for instance, in the mass bin $M_*\sim 10^{11.5}-10^{11.55}\,M_\odot$ at $z\sim 3.8-4$.
The very few galaxies on this tile are galaxies that belonged to the mass bin $M_*\sim 10^{11.3}-10^{11.35}\,M_\odot$ at the previous timestep
and that have experienced a sudden mass increase of $\sim 60\%$ due to a major merger (Fig.~\ref{fig1new}, middle panel).
However, most of the galaxies with $M_*\sim 10^{11.3}-10^{11.35}\,M_\odot$ at $z\sim 4-4.2$
do not experience a merger at the next timestep (Fig.~\ref{fig1new}, left panel).

Let us now compare the middle panel (the progenitors of all galaxies) to the right panel (the progenitors of red galaxies).
At high $M_*$, we hardly see any significant difference.
This should not surprise us because, at $z=0$, most galaxies with $M_*>10^{11}\,M_\odot$ are red.
At $M_*<10^{10.8}\,M_\odot$, we see that most red galaxies (right panel) have not grown in mass for the last half the cosmic lifetime (since $z\sim 1$).
In contrast, the overall galaxy population (middle panel) is dominated by objects that are still growing or that have grown until very recently.
The same difference is also seen when we compare the left panel and the middle panel.

While we have spent three paragraphs to clarify these differences,
the main result that emerges from a comparison of the right and the left panel in Fig.~\ref{fig1new} is, in fact, their similarity, particularly at high masses,
which constitute the focus of this article.
Hence, galaxy histories can be described as one-parameter family of curves, with the stellar mass at any given redshift as the key quantity that determines on which evolutionary tracks a galaxy lies.
This is also the reason why we can establish a one-to-one relation between $z=0$ galaxies and their high-$z$ parent population, at least in a statistical sense
\footnote{The assumption that galaxy star-formation and stellar masses form  a one-parameter family has been cited before in other contexts.  It is implicit, for example, in the star-forming main sequence of \citet{noeske_etal07} and \citet{noeske_etal07b}, in which galaxies of a given stellar mass today were fitted to a unique star-formation history.  It is also implicit in the halo abundance matching model of \citet{conroy_wechsler09}, in which they assumed that the star formation rate was uniquely determined by halo mass at each redshift.  These and similar prescriptions tend to produce a one-parameter family of galaxies labeled by mass.}.

\begin{figure*}
\noindent
\includegraphics[width=0.9\hsize]{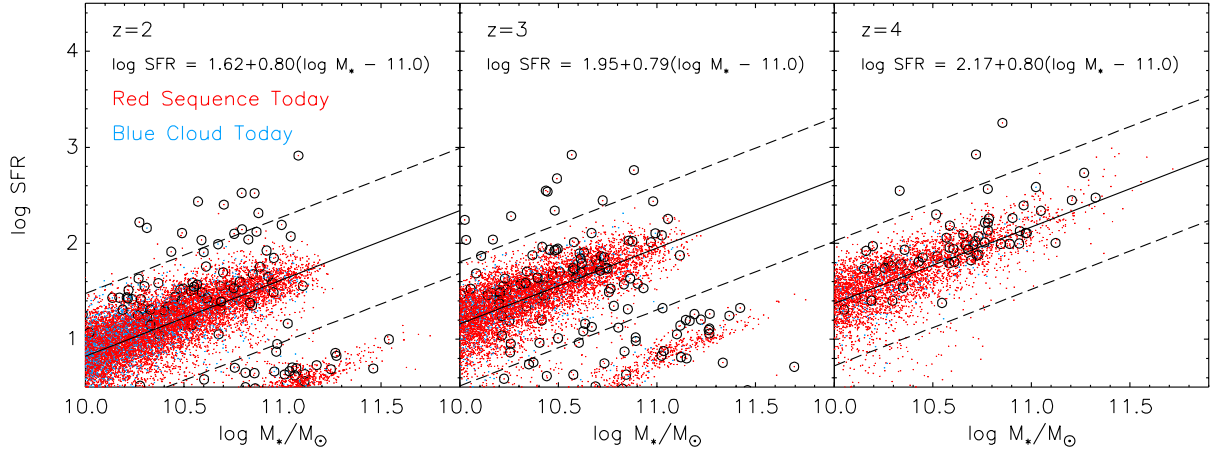}
\caption{SFR vs. galaxy stellar mass for all model galaxies in the
  computational box at $z=2$ (left), $z=3$ (centre) and $z=4$ (right).
  Each galaxy is shown with a red or a blue symbol depending whether
  its descendant at $z=0$ lies on the red sequence or the blue cloud.
   Symbols surrounded by a black circle correspond to ongoing
  mergers with mass ratio greater than 1:4 (major mergers). The solid lines show, for each redshift, the best linear least-square fit to the main sequence of SFGs, while the dashed lines mark $\pm3\sigma$.}
\label{fig1}
\end{figure*}
\begin{figure*}
\noindent
\includegraphics[width=0.9\hsize]{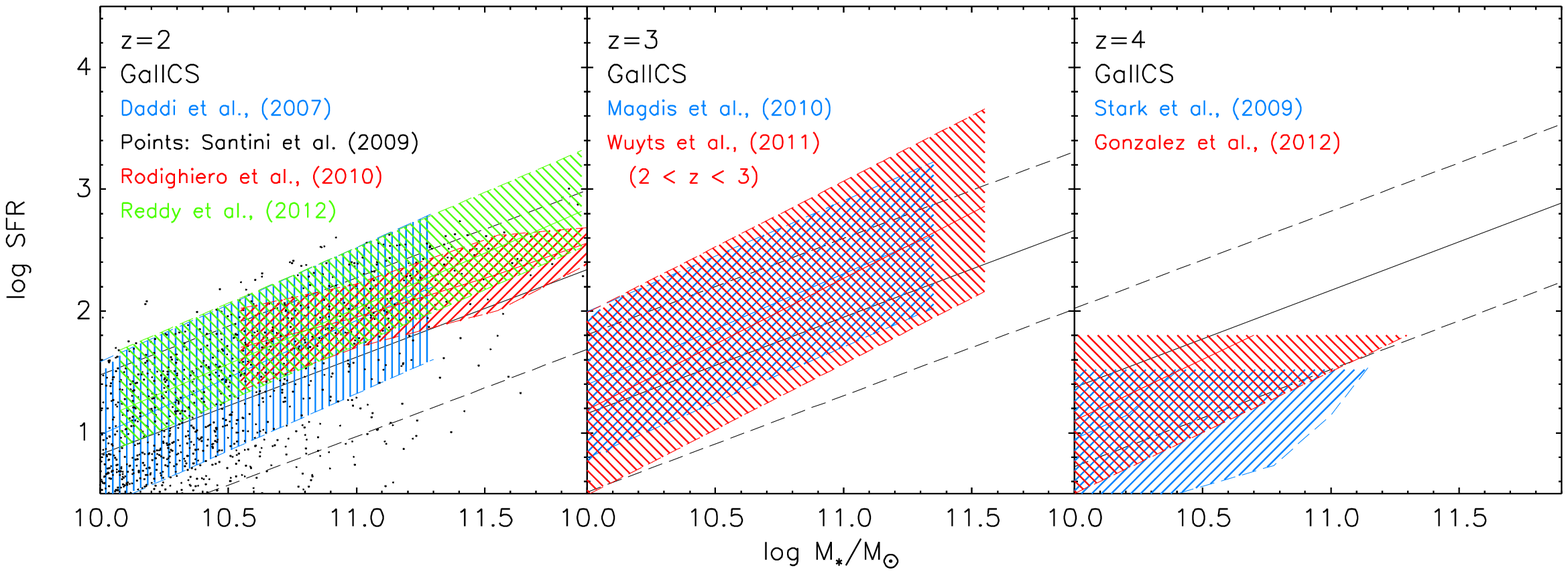}
\caption{The model results shown in Fig.~\ref{fig1} are compared with the galaxies of Santini et al. (2009; $z=2$) and different observational
determinations of the SFR-$M_*$ relation at $z=2$ \citep{daddi_etal07,rodighiero_etal10,reddy_etal12}, $z=3$ \citep{magdis_etal10,wuyts_etal11}, and
$z=4$ \citep{stark_etal09,gonzalez_etal12}.
The solid and dotted lines at $z=2\,3,\,4$ show the model relation and are the same as in Fig.~\ref{fig1}. They have been added to facilitate the comparison
with Fig.~\ref{fig1}.}
\label{sfr_mass_data}
\end{figure*}

This kind of diagram is useful to study also the evolution in the star
formation properties of the galaxy population.  The tracks in
Fig.~\ref{fig1bisnew} are the same backward evolutionary tracks that
we already showed in the {\it right} panel of Fig.~\ref{fig1new}, but now we have added
colour to display the median star formation rate (SFR; left panel), the median specific star formation rate (SSFR; middle panel), and the fraction
of red galaxies (right panel) on each mosaic tile. It is important to keep in mind that the tracks on the three panels of Fig.~\ref{fig1bisnew}
are all and only for galaxies that end up on the red sequence at $z=0$. This is why, in the right panel of Fig.~\ref{fig1bisnew}, the fraction of red galaxies at
$z=0$ is unity for all values of stellar mass.

Here a red galaxy is
  defined to be in the red sequence of the colour-mass diagram. We
  chose the division between the red sequence and blue cloud by eye
as the criterion to separate red and blue galaxies  at each
 timestep.

The left panel of Fig.~\ref{fig1bisnew} shows that, in our model, there is an almost one-to-one relation between the passive population of $z\sim 0$ massive galaxies (the galaxies with the lowest SFRs ever) and the population of massive $z=3-6$ galaxies with
SFR $>100\,M_\odot{\rm\,yr}^{-1}$ (which comprises the strongest starbursts in the Universe).
While at low-$z$, the most massive galaxies are those with the lowest SFRs,  at $z>3$ the highest SFRs are predicted to be in the most massive galaxies at those redshifts.

The right panel of Fig.~\ref{fig1bisnew} shows that, in our model, the
most massive galaxies grow along the blue sequence until $z\sim 3.5$,
and that the red sequence begins to emerge at $z\sim 3-3.5$.  After
this point, most giant ellipticals evolve passively.  One can see
clearly from Fig.~\ref{fig1bisnew} that the point where the SFR drops
and galaxy move to the red sequence is exactly where the curves
$M_*(z)$ go horizontal.  This is logical, since the stellar mass stops
increasing when star formation shuts down.  This shut down is a
  direct consequence of the assumption that gas accretion and star
  formation shut down above a critical mass $M_{\rm crit}(z)$, which
  underpins all the results of \citet{cattaneo_etal06},
  \citet{cattaneo_etal08}, and this article.  Restart of galaxy
growth at $z\lsim 0.5$ is due to dry mergers at the centres of groups
and clusters.  This effect is likely to be overestimated because our
model does not include tidal stripping (discussion in
\citealp{cattaneo_etal08}).

The middle panel of Fig.~\ref{fig1bisnew} shows a trend, in which the typical SSFR decreases with time and lower at higher stellar masses.


\section{High-redshift star-formers and their descendants}

Having described how we map the high-$z$ galaxy population into the low-$z$ one and vice versa, we are now
ready to present our results.  In this Section, we look at high-$z$
($z\sim 2-4$) SFGs and we follow their descendants at $z=0$.

Fig.~\ref{fig1} shows the SFR - stellar mass relation at $z=2,\,3,\,4$.  Most
galaxies (symbols) lie on a diagonal strip where the SFR grows with
the stellar mass $M_*$.  This strip is the main sequence of SFGs
\citep{noeske_etal07b,elbaz_etal11}.  Between $z=4$ and $z=3$, a second strip
emerges parallel to the former, the tip of which is visible in the panel at $z=3$.
It is the sequence of passive
galaxies.  Galaxies on the latter have much lower SFR for a given
stellar mass and this is where, already at $z= 3$, the most massive
galaxies lie.

We define the SFG population as follows.  We separate the main 
  sequence of SFGs from the passive population with the line $\log{\rm\,SFR} = 
 \log M_*+c(z)$ where $c(z) = (-10,-9.7,-9.5)$ for $z=(2,3,4)$.  We then
  perform a linear least squares fit to the galaxies that lie above
  this dividing line and that are more massive than
  $10^{10}M_\odot$.  {The fit is the solid line in Fig.~\ref{fig1}.}
  We compute the standard deviation $\sigma$ of the points from the
  fit and {consider the galaxies within 3$\sigma$ of the fit to be the
  main sequence of SFGs (the galaxy population within the two dashed lines).  We define an SFG to
    be any galaxy above the lower dashed line, including the outliers
    from the main sequence of SFGs.}
  We, therefore,
do not adopt an absolute SFR criterion to define SFGs, though we
introduce a stellar-mass cut at $M_* >10^{10}\,M_\odot$.  {One can
  see from Fig.~\ref{fig1}} that, with this cut, our simulated-galaxy sample is
`complete' down to $\sim (100,60,30) \,M_\odot{\rm\,yr}^{-1}$ at
$z=(4,3,2)$.  In the next Section, we shall start from massive red
galaxies at $z=0$ and we shall track their progenitors back in time.

In our model, the normalisation of the SFR - $M_*$ relation decreases from high to
low $z$ for both star-forming and passive galaxies.  A least squares
fit to {the} main sequence of SFGs, ${\rm
    Log\,SFR}/(M_\odot{\rm\,yr}^{-1})=a({\rm\,Log\,}M_* /M_\odot -
  11) +b$} yields similar slope of $a\sim 0.80$ in each
redshift bin, but the normalisation at $M_*=10^{11}\msun$ 
  decreases from ${\rm SFR}\sim 160\,M_\odot{\rm\,yr}^{-1}$ at $z=4$ to 
  ${\rm SFR}\sim 100\,M_\odot{\rm\,yr}^{-1}$ at $z=3$ and
  ${\rm SFR}\sim 40\,M_\odot{\rm\,yr}^{-1}$ at $z=2$ 
Hence, the
number of galaxies above a given SFR decreases with time.  For
instance, the number density of galaxies with {SFR
  $>100\,M_\odot{\rm\,yr}^{-1}$} is predicted to decrease from
{$\sim 170$ per $(150\,\mpc)^{3}$ at $z=4$ to $\sim 50$ and $\sim 7$
  per $(150\,\mpc)^{3}$} at $z=3$ and $z=2$, respectively.  
  
 Fig.~\ref{sfr_mass_data} compares the model results of Fig.~\ref{fig1} 
 with observational determinations of the SFR - $M_*$ relation.
 Daddi et al. (2007; $z=2$), Santini et al. (2009; $z=2$), Rodighiero et al. (2010; $z=2$), Reddy et al. (2012; $z=2$), Stark et al. (2009; $z=4$), and
 Gonzalez et al. (2012; $z=4$ used 
 the Salpeter IMF in their estimates of $M_*$,
while Magdis et al. (2010; $z=3$) and Wuyts et al. (2011; $z=3$) 
 used the Chabrier IMF.  
 We have
   scaled their values to be consistent with the Kennicutt IMF used in \citet{cattaneo_etal06} and \citet{cattaneo_etal08}.
To pass from the Salpeter IMF and the Chabrier IMF to the Kennicutt IMF,
we have used the relations found in \citet{bell_etal03,fardal_etal07,treyer_etal07}.

The slope and the scatter of the model relations at $z=2$ and $z=3$ are quite similar to those that we find in the data.   
 However, the normalisation of the SFR - $M_*$ relation at $M_*=10^{11}\msun$  that we find in our model at $z=2$
 is $\sim 2-3$ times lower than that found in observational studies  
 \citep{daddi_etal07,santini_etal09,rodighiero_etal10,reddy_etal12}. 
We find the same discrepancy at $z=3$, when we compare our results to the observational determinations by \citet{magdis_etal10} and \citet{wuyts_etal11}.

At $z=4$ our SFR - $M_*$ relation appears to be higher than the observational determinations by \citet{stark_etal09} and \citet{gonzalez_etal12}.
However, the comparison with the data at $z=4$ requires extreme caution for two reasons.
   First, the blue and the red hatched ares finish abruptly at $\sim 10^{1.5}\,M_\odot/$yr and  $\sim 10^{1.8}\,M_\odot/$yr, respectively,
   because the relation at $z=4$ is given in bins of SFR, rather than in bins of $M_*$.
   Secondly, the data by \citet{stark_etal09} are not dust corrected.

Globally, the model is not doing too bad considering that the rough agreement in Fig.~\ref{sfr_mass_data} has been obtained without tuning any free 
parameter.
We have simply taken a model that is already published \citep{cattaneo_etal06,cattaneo_etal08} and that works well at $z=0$, and we have analysed its predictions
 for the SFR - $M_*$ relation at $z=2-4$.
 Admittedly we have also required that this model reproduces the luminosity function of Lyman-break galaxies at 1700 angstrom restframe.
 However, this is no guarantee to reproduce the SFR - $M_*$ relation, particularly since many SFR measurements come from infrared data.

In closer detail,} the discrepancy with the observed SFR - $M_*$ relation at $z=2$ was not unexpected because we know that the cosmic SFR density predicted by our model at $z=2$ is at the lower limit of
 the range allowed by observations (Fig.~8 of Cattaneo et al. 2006), and that all semianalytic models underpredict sub-mm counts,
 unless they invoke a top-heavy stellar initial mass function \citep{baugh_etal05,lacey_etal08}.
 
 We should also note,  however, that at $z=1$, where there are more data, \citet{noeske_etal07b} and \citet{chen_etal09} find a substantially 	lower
 normalisation of  the SFR - $M_*$ relation than \citet{elbaz_etal07} and Santini et al. do (Elbaz et al. 2007 and Daddi et al. 2007 are the same group).
 
 \begin{figure}
\noindent
\includegraphics[width=0.9\hsize]{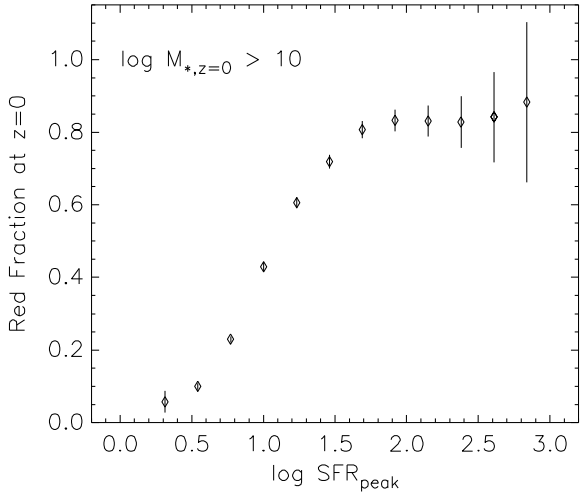}
\caption{Fraction of galaxies on the red sequence at $z=0$ as a
  function of peak star formation rate. The error bars are Poissonian
  errors.  This figure {includes} all galaxies with
  $M_*>10^{10}\,M_\odot$ at $z=0$.}
\label{fig1bis}
\end{figure}
  
Having discussed the comparison with the data, we now take a closer look at the results of our model.

  Fig.~3 shows
that, at $z=4$, a lot of the SFGs with
SFR$>100\,M_\odot{\rm\,yr}^{-1}$ are objects at the massive end of the
SFR sequence.  In contrast, at $z=2$, SFGs with
SFR$>100\,M_\odot{\rm\,yr}^{-1}$ are outliers from the main sequence.

The main sequence of SFGs represents the normal SFG population at a given
redshift.  Yet, at $z\sim 4$, the tip of the main sequence of SFGs reaches SFRs
of about $1000\,M_\odot{\rm\,yr}^{-1}$.

Some galaxies have very high SFRs (up to thousands of Solar masses per
year) that cause them to be outliers from the main sequence of SFGs.
It is apparent from Fig.~\ref{fig1} that these galaxies are almost
always mergers, because objects which have experienced a greater
  than 1:4 merger in the last 300$\,$Myr have been surrounded with a
  black circle.

Fig.~\ref{fig1} suggests that mergers are necessary to explain the highest SFRs
at any given $z$, but that they do not drive the bulk of star
formation at high redshift, most of which occurs along the main
sequence of SFGs.  The circled symbols account for {0.6, 1.2, and 1.4$\%$}
of the objects on the main sequence of  SFGs  {(i.e., those between the $3\sigma$ lines in Fig.~\ref{fig1})}
at $z=2,\,3,\,4$, respectively.  The percentages increase
to {54, 92, and 100$\%$} if we look at outliers above the main
sequence of SFGs, defined as galaxies that lie more than
{$3\sigma$} above the fit to the main sequence of SFGs (upper line in
Fig.~\ref{fig1}).

\begin{figure*}
\noindent
\includegraphics[width=0.9\hsize]{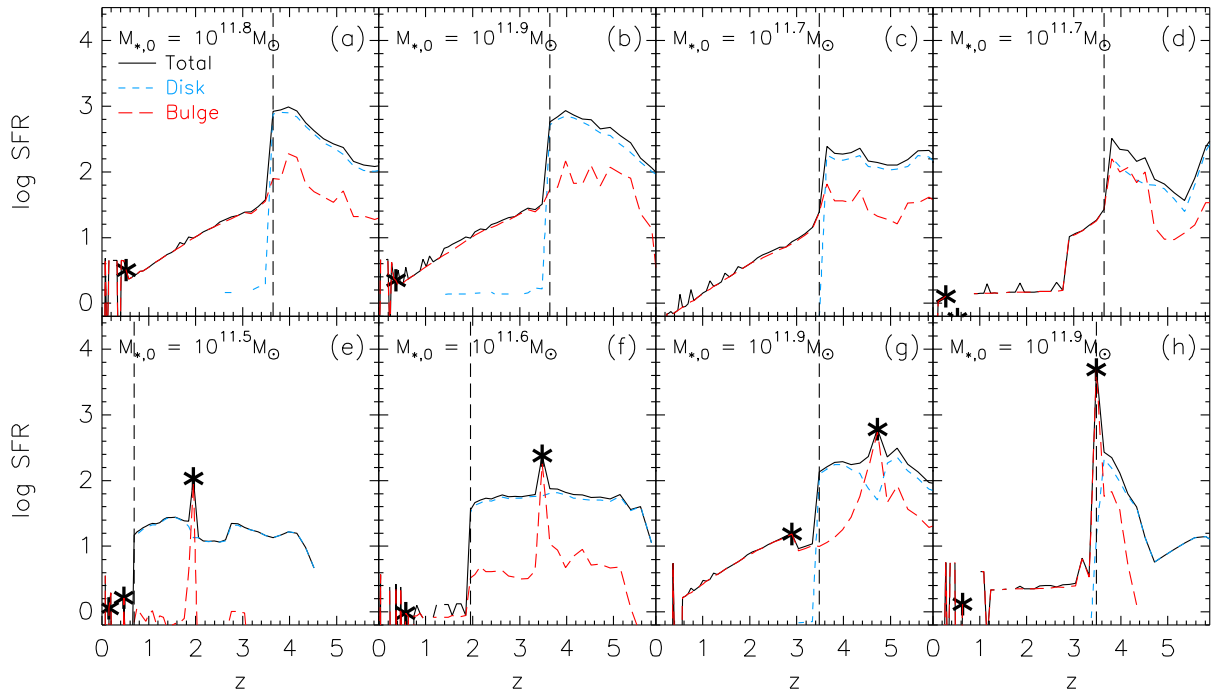}
\caption{The SFR histories of eight red-sequence galaxies selected for
  illustrative purposes. Each SFR history has been decomposed into
  disc (blue) and bulge (red) star formation. The stellar mass of each
  galaxy at $z=0$ has been shown in the corresponding panel.  Major
  mergers (with ratio $>1:4$) have been marked with {asterisks}. Most of them are
  `dry' (dissipationless) mergers occurring after the galaxies have
  ceased to make stars. `Wet' (gas-rich) mergers are accompanied by
  strong starbursts, characterised by a sharp rise and a fall in the
  bulge SFR.  The fact that the bulge SFR declines continuously and is
  higher than the disc SFR after quenching is an artifact due to
  assuming a gas-surface-density threshold for disc star formation but
  not for bulge star formation, which, after shutdown, is fed by
  stellar mass loss only. }
\label{fig2}
\end{figure*}

Galaxies have been plotted in red or blue depending on whether their
descendants lie on the red sequence or the blue cloud of the colour -
$M_*$ diagram at $z = 0$. The criterion used to separate red and blue
galaxies is {$U-B > 0.95 +0.06(\log M_*/\msun - 10.0)$}.

{A strong result that emerges from Fig.~3 is that virtually all
  massive {SFG} galaxies evolve into the red sequence at $=0$.  The
  fraction of SFGs that end up on the red sequence depends slightly on
  mass, SFR and redshift.  We find that $\sim 95 - 98\%$ of high-$z$ SFGs
  with {$M_*>10^{11}\,M_\odot$} evolve into red-sequence galaxies by
  $z=0$ (these values correspond to $z=2$ and $z=4$, respectively).
  At lower masses ($10^{10}\,M_\odot<M_*<10^{11}\,M_\odot$), the
  fraction of SFGs that end up of the red sequence is lower and
  exhibits a slightly stronger dependence on redshift.  The SFGs with $10^{10}\,M_\odot<M_*<10^{11}\,M_\odot$
  that evolve into red-sequence galaxies by $z=0$ are
  86$\%$, 91$\%$, and 97$\%$ at $z=2,\,3,\,4$, respectively.}

Above a given SFR, say $100\,M_\odot{\rm\,yr}^{-1}$, objects
    have a substantial probability of ending up on the red sequence at
    $z=0$ (96$\%$, 95$\%$ and 97$\%$ for objects with SFR$>100\,M_\odot{\rm\,yr}^{-1}$
    at $z=2,\,3,\,4$, respectively).

  Fig.~5 shows the probability that a $z = 0$ galaxy is on the red
  sequence as a function of its peak SFR.  It shows that 80\% of all
  objects that have ever reached $> 40 \msun{\rm yr}^{-1}$ evolve into
  red-sequence galaxies.
  
  We do not argue that the peak SFR of a galaxy is the most important factor in determining if it ends up on the red sequence or the blue cloud.
  But since the SFR of a galaxy at any time put a lower limit to its peak SFR,
  Fig.~4 is useful because, given a SFR measurement, it gives the minimum probability that an observed galaxy has to end up on the red sequence
  at $z=0$.

In fact, we do not argue at all that absolute SFR provides  a physical criterion 
to tell if a galaxy becomes quenched.
Galaxy mass (due to its link to halo mass) and morphology are, in this sense, much more relevant.
The purpose of this analysis is rather to show the correlation between one of the most basic properties that one can measure in 
a high-$z$ galaxy (SFR) and what our model predicts for its later evolution.

\section{The progenitors of massive red galaxies}

In Section~3, we have looked at high-redshift ($z\sim 2-4$) SFGs with
$M_*>10^{10}\,M_\odot$ and we have found that most of them evolve into
massive red-sequence galaxies by $z=0$.
We now do the opposite. We start from massive red-sequence galaxies in
the local Universe and we reconstruct their SFR histories by tracking
their most massive progenitor back in time.

Fig.~\ref{fig2} has been constructed by selecting eight galaxies with $M_*\gsim
10^{11.5}\,M_\odot$ today.  These galaxies have been chosen to illustrate the variety of star formation histories of today's massive red galaxies The total SFR as a function of $z$ (black
solid line) has been decomposed into the contributions of disc and
bulge.  Mergers with mass ratios greater than 1:4 (major mergers) are
highlighted with {an asterisk} symbol.  The vertical dashed lines
show the redshifts at which the dark matter haloes of these galaxies
have passed $M_{\rm crit}$.

In galaxies (a), (b) and (c), disc star formation dominates the total
SFR before star formation is quenched at $3<z<4$ (the fact that the
bulge SFR is higher than the disc SFR after quenching is an artifact
due to assuming a gas-surface-density threshold for disc star
formation but not for bulge star formation).  In galaxy (d), the two
are comparable just before star formation is quenched.  Galaxy (c)
experienced no major mergers
(in Fig.~\ref{fig2}, major mergers are marked with asterisks). Galaxies (a), (b) and (d) experienced
one major merger each at $z<1$, long after they ceased to make stars.
In conclusion, the mergers experienced by galaxies (a), (b), (c) and
(d) are dissipationless and the bulge star formation in galaxies (a),
(b), (c) and (d) is driven not by mergers but rather by disc
instabilities.

Galaxies (e), (f), (g) and (h) had 2-3 major mergers each and they
exhibit strong SFR peaks at the time of their first merger, which
occurs before quenching, when the galaxies were still gas-rich.  There
is a clear direct link between dissipative mergers and strong starbursts in the bulge component,
which cause it to dominate the SFR temporarily.  In galaxies (e) and (f), mergers cause the SFR to
increase by a factor of $\sim 4$. In galaxy (h), the SFR increases by
a factor of $\sim 10$.  It is also interesting to note that while
galaxies (e), (f) and (g) go back to normal `quiescent' star formation
after a merger-driven starburst - if one can call quiescent a
galaxy like (g) with a disc SFR $>100\,M_\odot{\rm\,yr}^{-1}$), - in
galaxy (h) the merger-driven starburst coincides with the
shutdown of star formation.

Merger-driven starbursts are, by construction, short-lived.  We have
verified that in galaxies (e), (f) and (g) only a small fraction of
the final galaxy stellar mass was formed in the bulge even if the
highest SFR occurred in the bulge.  Galaxy (h) is the prototypical
example of an elliptical galaxy that makes most of its stars in a
single merger-driven burst at high $z$. However, objects of this type are
exceedingly rare.  They can be counted on the fingers of one hand in
our computational volume of $(150{\rm\,Mpc})^3$.
 
With insight from these particular cases, we are now ready to look at
the statistical properties of the progenitors of red-sequence
galaxies.
We split the latter into three ranges of stellar mass at $z=0$ (lower intermediate mass: $10^{10.5}M_\odot<M_*<10^{11}M_\odot$; upper
intermediate mass: $10^{11}M_\odot<M_*<10^{11.5}M_\odot$; high mass:
$M_*>10^{11.5}M_\odot$) and characterize the SFR histories of the disc
and the bulge of each galaxy in terms of two properties: the peak SFR
and the redshift at which the SFR reaches its peak value. 

\begin{figure*}
\noindent
\includegraphics[width=0.6\hsize]{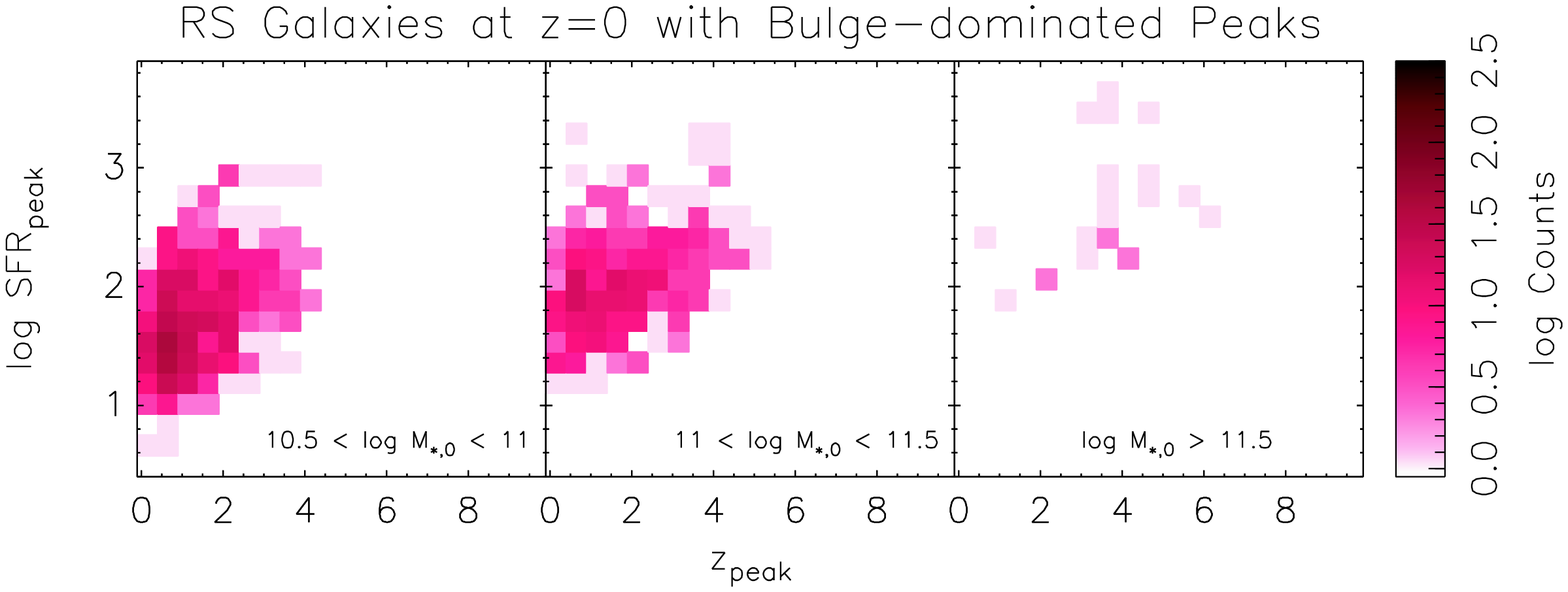}
\includegraphics[width=0.6\hsize]{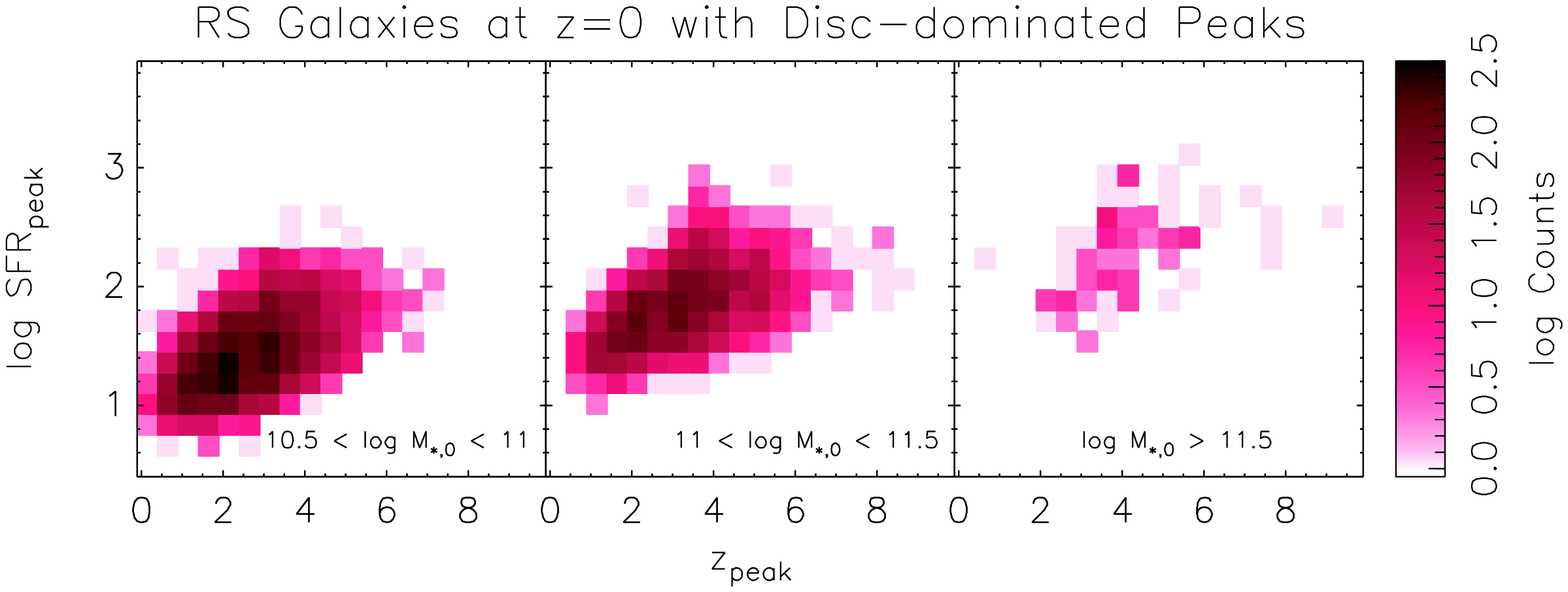}
\caption{The joint distribution of peak SFR and redshift of peak SFR for $z=0$ red-sequence galaxies in three bins of $z=0$ stellar mass: $10^{10.5}M_\odot<M_*<10^{11}M_\odot$, 
$10^{11}M_\odot<M_*<10^{11.5}M_\odot$, and $M_*>10^{11.5}M_\odot$.
The upper and lower panels show the SFR$_{\rm peak}$ - $z_{\rm peak}$ distribution for galaxies with 
${\rm SFR}^{\rm bulge}_{z=z_{\rm peak}}>{\rm SFR}^{\rm disc}_{z=z_{\rm peak}}$ and ${\rm SFR}^{\rm bulge}_{z=z_{\rm peak}}<{\rm SFR}^{\rm disc}_{z=z_{\rm peak}}$, respectively.
}
\label{fig3_0}
\end{figure*}

\begin{figure*}
\noindent
\includegraphics[width=0.6\hsize]{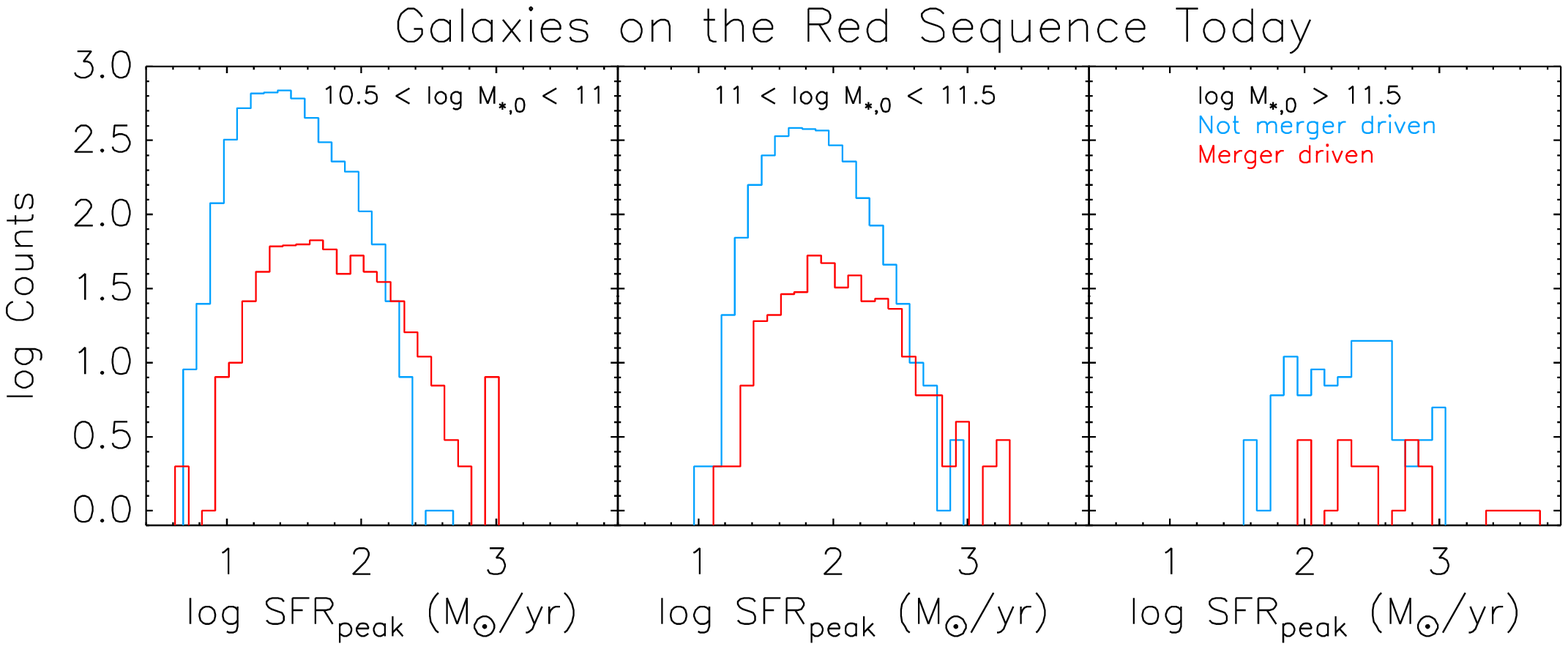}
\caption{Distribution for the value of the peak SFR for red-sequence
  galaxies in three intervals of stellar mass at $z=0$:
  $10^{10.5}\,M_\odot<M_*<10^{11}\,M_\odot$ (left),
  $10^{11}\,M_\odot<M_*<10^{11.5}\,M_\odot$ (centre) and
  $10^{11.5}\,M_\odot<M_*<10^{12}\,M_\odot$ (right).  The distribution
  of SFR$_{\rm peak}$ has been shown separately for
  {disc-dominated peaks} (blue histograms) and
  {bulge-dominated peaks} (red histograms).}
\label{fig3}
\end{figure*}

\begin{figure*}
\noindent
\includegraphics[width=0.6\hsize]{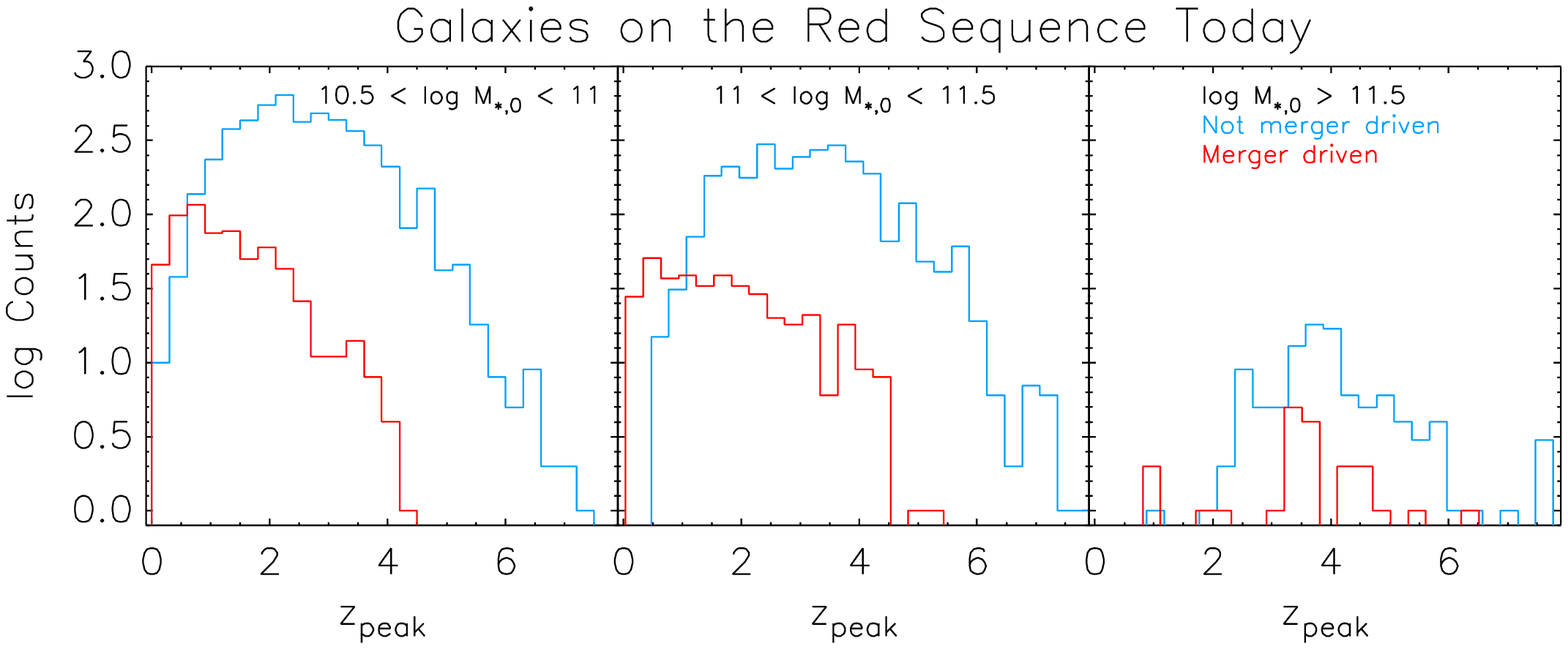}
\caption{Distribution for the value of the redshift $z_{\rm peak}$ at
  which the SFR {reaches its maximum} for red-sequence galaxies in three
  intervals of stellar mass at $z=0$:
  $10^{10.5}\,M_\odot<M_*<10^{11}\,M_\odot$ (left),
  $10^{11}\,M_\odot<M_*<10^{11.5}\,M_\odot$ (centre) and
  $10^{11.5}\,M_\odot<M_*<10^{12}\,M_\odot$ (right).  The distribution
  of $z_{\rm peak}$ has been shown separately for 
  {disc-dominated peaks} (blue histograms) and
  {bulge-dominated peaks} (red histograms).}
\label{fig4}
\end{figure*}

Fig.~\ref{fig3_0} shows the joint distribution of SFR$_{\rm peak}$ and
$z_{\rm peak}$ for galaxies in which in the peak SFR is linked
{to} star formation in the bulge component (above) and {to}
the disc component (below).  The histograms in
Figs.~\ref{fig3}-\ref{fig4} show the distributions for SFR$_{\rm
  peak}$ and $z_{\rm peak}$ separately.  {In these figures, blue
  histograms are those galaxies whose peak SFR was in the disc (they
  correspond to the {lower} panels of Fig.~\ref{fig3_0}), while red
  histograms are those whose peak SFR was in the bulge (they
  correspond to the {upper} panels of Fig.~\ref{fig3_0}).}

\begin{figure*}
\noindent
\includegraphics[width=0.6\hsize]{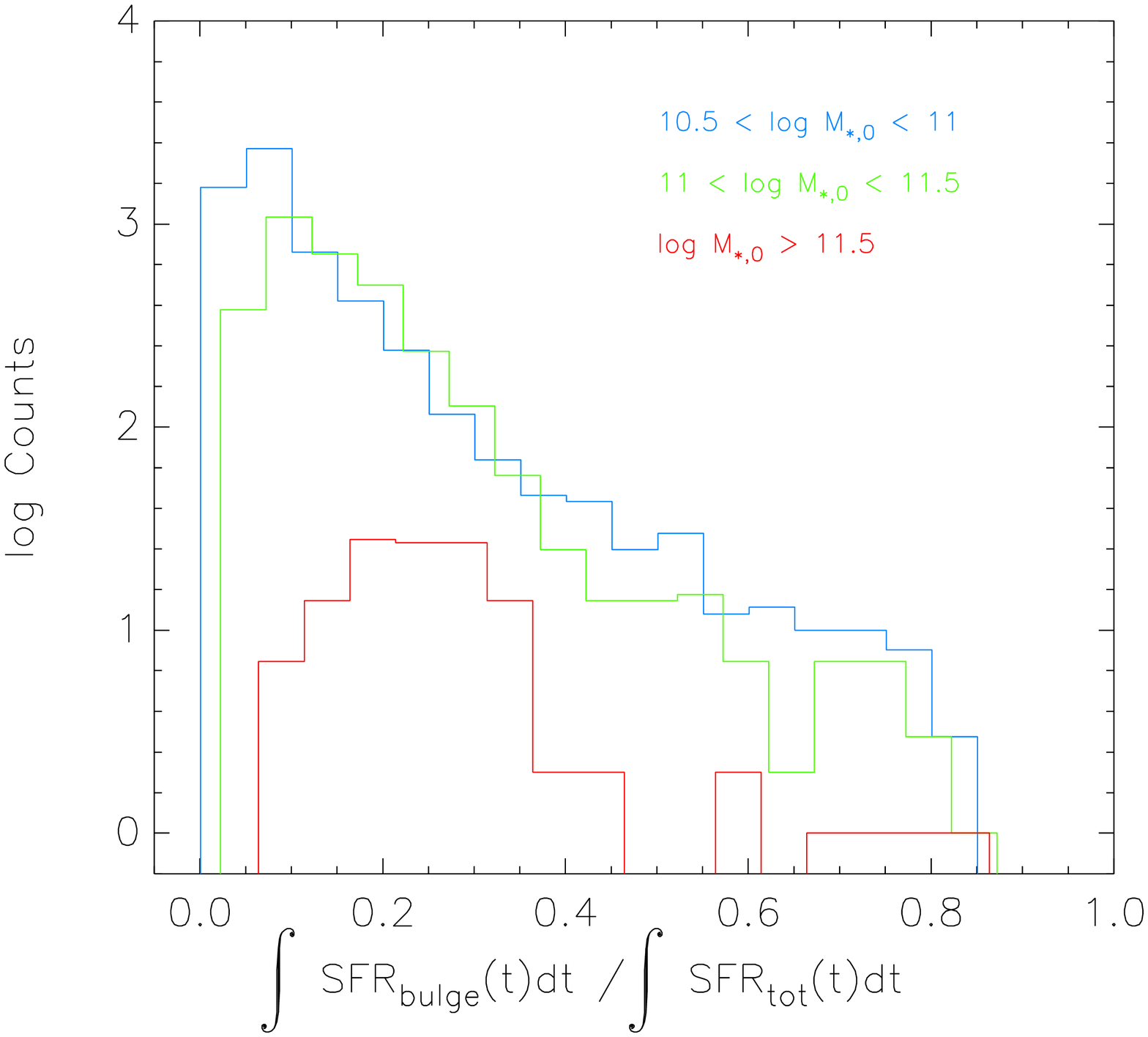}
\caption{The contribution of bulge star formation to the total in situ star formation of red sequence galaxies (computed by following the galaxies' main progenitors).
The distribution of values for $\int_0^{t_0}{\rm SFR_{bulge}(t)\,d}t/\int_0^{t_0}{\rm SFR_{tot}(t)\,d}t$ is shown for three bins of stellar mass at $z=0$:
$10^{10.5}\,M_\odot<M_*<10^{11}\,M_\odot$, $10^{11}\,M_\odot<M_*<10^{11.5}\,M_\odot$, and $M_*>10^{11.5}\,M_\odot$.}
\label{integralsofbulgeSF}
\end{figure*}

Figs.~\ref{fig3_0}-\ref{fig4} show that the
typical peak SFR is tens of Solar masses per year for lower
intermediate-mass red galaxies, {about} a hundred Solar masses per
year for upper intermediate-mass red galaxies and {several} hundreds
Solar masses per year for high-mass galaxies, but this is only a
statistical statement. Even among lower intermediate-mass galaxies,
there are objects with SFR$_{\rm peak}\sim
1000\,M_\odot{\rm\,yr}^{-1}$. In all ranges of mass, galaxies in which
the peak SFR occurs {\it in the disc} are much more numerous than
galaxies in which the peak SFR occurs in the bulge.  To put it
quantitatively, only $11\%$, $12\%$, and $16\%$ of the red-sequence
galaxies in the stellar mass intervals
$10^{10.5}\,M_\odot<M_*<10^{11}\,M_\odot$,
$10^{11}\,M_\odot<M_*<10^{11.5}\,M_\odot$, and
$10^{11.5}\,M_\odot<M_*<10^{12}\,M_\odot$ have experience{d} a
SFR peak dominated by bulge star formation.

As we argued at the end of Section~2.3, disc star formation is always
stream-fed in the sense that, by construction, discs can acquire gas
from smooth accretion only.  In contrast, both mergers and rapid
smooth gas accretion contribute to bulge star formation. Major mergers
do it directly, by bringing gas into the starburst/bulge
component. Gas accretion and minor mergers do it indirectly, by
causing a sudden increase in the gas content of the disc, which
becomes unstable and transfers mass to the bulge.  Thus the
{counts in the upper panel of Fig.~\ref{fig3_0} and the} red
histograms in Figs.~\ref{fig3}-\ref{fig4} represent {\it upper limits}
to merger-driven SFR, and we refer to these histograms loosely as the
merger-driven contribution.
 
Therefore, our model makes two strong predictions.  The first is that,
in most red galaxies, star formation was stream-fed, not
merger-driven.  The second is that, conversely, mergers are
responsible for the most intense episodes of star formation at each
redshift {see Fig.~\ref{fig3}}, in agreement with data over a
broad range of redshifts from $z\sim 0$ to $z\sim 2$.

The first conclusion can also be verified directly by taking the total SFR history
and the bulge SFR history of each galaxy, computed by tracking the main progenitor back in time
(i.e. the equivalent of the black and the red curves in Fig.~\ref{fig2}), and
by integrating them over time, to determine what fraction of the star formation that has occurred has been bulge star formation.
Fig.~\ref{integralsofbulgeSF} shows the distribution of values for $\int_0^{t_0}{\rm SFR_{bulge}(t)\,d}t/\int_0^{t_0}{\rm SFR_{tot}(t)\,d}t$ in the three
bins of galaxy stellar mass $10^{10.5}\,M_\odot<M_*<10^{11}\,M_\odot$, $10^{11}\,M_\odot<M_*<10^{11.5}\,M_\odot$, $M_*>10^{11.5}\,M_\odot$
(stellar masses at $z=0$; $t_0$ is the current age of the  Universe; and this figure is only for red-sequence galaxies).
The average contribution of merger driven star formation for the three stellar mass bins is $11\%$, $15\%$, and $26\%$, respectively.
Galaxies that formed most of their stars in merger-driven starbursts are rare, though it is quite usual that a quarter of the in-situ star formation
in a massive early-type galaxy occurs via this mode. Most of the stars in a giant elliptical were not formed in situ, i.e. they were formed in smaller objects that
merged with the galaxy, so we do not include either in the $\sim 1/4$ of bulge star formation or the $\sim 3/4$ of disc star formation.

Fig.~\ref{fig4} shows the distribution for the redshift $z_{\rm peak}$ at which
the peak SFR occurs for the three ranges of masses.  For lower
intermediate-mass red galaxies{,} $z_{\rm peak}\sim 2$. For upper
intermediate-mass red galaxies{,} $z_{\rm peak}\sim 2.5-3.5$. For
high-mass red galaxies{,} $z_{\rm peak}\sim 4$.

The decrease of $z_{\rm peak}$ at lower masses is an aspect of
downsizing.  In our model, downsizing occurs because haloes with a
lower mass at {fixed} $z$ cross $M_{\rm crit}$ at a lower redshift
(Fig.~4 of Cattaneo et al. 2008).  In the same way $M_{\rm crit}$ is
constant at $z\lsim 3$ and increases at higher $z$, this also applies
to the characteristic stellar mass $M_*^{\rm crit}$ with which
galaxies enter the red sequence.  The characteristic star formation
rate of a galaxy that enters the red sequence at cosmic time $t$ is
{SFR $\sim M_{\rm crit}(t)/t$}, which must
decrease with $t$, since $\sim M_{\rm crit}(t)$ is a non-growing
function of $t$.

In \citet{cattaneo_etal08}, we analysed the time of peak SFR as a function of the final stellar mass for massive red-sequence galaxies.
The results were found to be in agreement with that  \citet{thomas_etal05} inferred from the spectra of local early-type galaxies
(Figs.~7-8 of Cattaneo et al. 2008).

Stream-fed star formation is responsible for most of the SFR peaks at
all redshifts except perhaps {at} low-redshift ($z<1$).  In the
stellar mass bins $10^{10.5}\,M_\odot<M_*<10^{11}\,M_\odot$,
$10^{11}\,M_\odot<M_*<10^{11.5}\,M_\odot$, and
$10^{11.5}\,M_\odot<M_*<10^{12}\,M_\odot$, the {red} galaxies with
$z_{\rm peak}<1$ are {$9\%$, $5\%$, and $2\%$}, respectively.  If
we look at this specific population, then the percentages of galaxies
with merger-driven SFR peaks rise to {$56\%$, $80\%$ and $67\%$} (instead
of $11\%$, $12\%$, and $16\%$, as we saw before).  These values would
have been even higher if we took $z<0.5$ instead of $z<1$ (we would have found
 81\%, 98\% and 100\%, respectively).
{Fig.~\ref{fig3_0} tells us that these merger-driven peaks are
  typically about $50\,M_{\odot} {\rm yr}^{-1}$ for the lower-intermediate
  mass bin, and about $100\,M_{\odot} {\rm yr}^{-1}$ for the upper
  intermediate mass bin.}  Merger-driven star formation is therefore
relevant to those $M_*\sim 10^{11}M_\odot$ ellipticals that have
formed via low-redshift gas-rich mergers such as those that we witness
in local ULIRGs (see \citealp{cattaneo_etal11} for a discussion of the
role of gas-rich mergers in the build-up of the galaxy population).
The same trend is also seen in massive red-sequence galaxies: the SFR
peak is usually stream-fed and is usually at $z\sim 4$ but in the few
objects with merger-driven SFR peaks, the SFR peak is normally around
$z\sim 3$.  Most of the merger activity of massive red galaxies is
predicted to be dissipationless and to occur at fairly low redshifts
\citep{cattaneo_etal08,delucia_blaizot07}.

 \begin{figure*}
\noindent
\includegraphics[width=0.98\hsize]{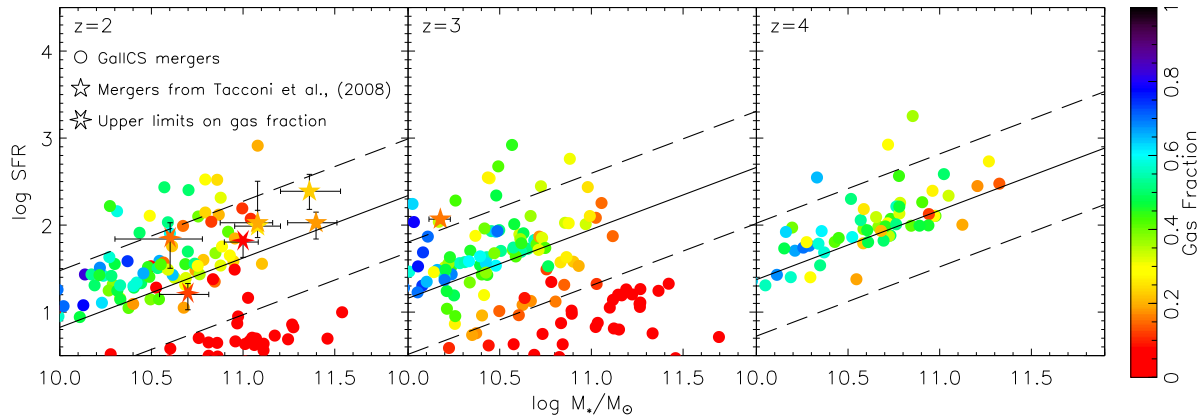}
\caption{The gas content of major mergers at $z=2\,3,\,4$.  GalICS
  galaxies that have experienced a merger in the last 300$\,$Myr are
  shown as circles on the SFR-$M_*$ diagram.  Mergers observed in
    Tacconi et al., (2008) are marked as stars (six-point stars correspond to upper limits on the gas fraction). Their stellar masses
    and SFRs have been adjusted
   to the
    Kennicutt IMF.  Each symbol is colour-coded according to the
  merger gas content, i.e., $M_{\rm gas}/(M_*+M_{\rm gas})$.  
  jwb{A
    circled star symbol indicates that the gas fraction is an upper
    limit, while the rest of the data from Tacconi et al. have
   uncertainties of about 0.08 in the fraction.}  
    The lines show the
  position of the main sequence of SF galaxies. They are the same as
  in Figs.~\ref{fig1}
  and~\ref{sfr_mass_data}.}
\label{merger_gasfraction}
\end{figure*}

\section{Discussion and conclusion}

To summarize, we find that: i) most galaxies that have
experienced SFRs $>40\,M_\odot{\rm\,yr}^{-1}$ evolve into red-sequence
galaxies, ii) {$80\%$} of the red-sequence galaxies with
$M_*>10^{11}M_\odot$ passed through a phase of high SFR (SFR$_{\rm
  peak}>40\,M_\odot{\rm\,yr}^{-1}$) at $z>1$ , and iii) in $\gsim 90\%$ of
these cases, this phase was stream-fed (merger ratio at peak SFR less
than 1:4).

Still, mergers are necessary to explain the galaxies with the highest
SFRs (e.g., the rare objects with SFRs of thousands of Solar masses
per year).  Furthermore, if we concentrate on $z=0$ red-sequence galaxies with peak SFR
at $z<1$ ({$9\%$, $5\%$, and $2\%$} of the objects in the three mass
intervals), then the percentages of merger-driven SFR peaks rise to
{$56\%$, $80\%$ and $67\%$}, respectively.  These values would have been
even higher if we took $z<0.5$ instead of $z<1$.  Hence, mergers are
important for that small percentage of latter-day ellipticals that
form in local ULIRGs, even though this cannot be the path via which most ellipticals formed \citep{ostriker80}.

These figures (and all the others in this article) should be
  taken with caution because our model contains many arbitrary
  assumptions.  Their uncertainties cannot be easily quantified but we
 believe they are probably greater than $10\%$.  
The purpose of the percentages quoted in this
  article is to illustrate the magnitude of the discussed phenomena. A
  change from $56\%$ to $67\%$ in the fraction of merger-driven peaks
  from one mass bin to another is not significant when one considers
  the uncertainties of the model.  
  However, we prefer to give the
  exact numbers returned by our model accompanied by this disclaimer
  than to use vague language, where any quantitative information is
  lost.

The accuracy of a semianalytic model is established {\it a posteriori} 
by how well it the reproduces the observations.
Our model is in very good agreement with local data.
The main problem is the discrepancy with
the normalisation of the SFR - $M_*$ relation at $z=2-3$, which we have discussed in Section~4.

At $M_*=10^{11}\msun$, our model predicts SFRs that are $\sim 2-3$ times lower than those inferred observationally by \citet{daddi_etal07},  \cite{santini_etal09}, \citet{rodighiero_etal10}, \citet{magdis_etal10}, \citet{wuyts_etal11}, \citet{reddy_etal12},
and,  at the massive end of the main sequence of SFGs, by \citet{huang_etal09}.
Our average SFR at a
given stellar mass would still be too low compared to these data even if all the accreted
gas were immediately converted into stars.
The same basic problem
 is found in the Munich model (e.g. \citealp{kitzbichler_white07})
  and in the Durham model, which proposes a top-heavy stellar initial 
  mass function (IMF) as a possible solution
(e.g. \citealp{baugh_etal05,lacey_etal08}).

Since \citet{noeske_etal07} and \citet{chen_etal09} find a lower
normalisation at $z=1$ than the authors mentioned above, it is
possible that these authors' SFR measurements at $z=2$ need to be
lowered, too, or that their masses are systematically underestimated.
However, we also know that semianalytic models underpredict sub-mm
counts, unless they invoke a top-heavy IMF.  So either a top-heavy IMF
{\it is} the solution, in which case there may be nothing wrong with
our predicted SFRs, and it is the observationally derived SFRs that
need to be corrected accordingly, or this signals a problem in our
galaxy formation model.  For instance, it is possible that our specific SFRs at $z=4$
are too high (Fig.~4) because gas accreted at $z\gsim 4$ is converted into
  stars too rapidly (see, e.g., \citealp{krumholz_dekel12}), and this
  may be the reason why, in the model, there is not enough star
  formation at later epochs ($z\sim 2-3$; Fig.~\ref{fig1}),
  though it also possible that 
$z=4$ SFRs inferred by \citet{stark_etal09}
are underestimated due to dust.

In fact, one should also remember that the observational determinations of the SFR-$M_*$ relation
are themselves model-dependent. Measured SFRs and stellar masses depend on stellar-population synthesis models, dust models, and assumptions about the
galaxies' SFR histories.

 The question here is: to what extent these open issues affect our confidence in our results?

Could {the underpredicted SFRs} for a given $M_*$ {be due to an
underestimate of} the importance of mergers?  Could our conclusion that
the bulk of star formation in the ancestors of red galaxies is
stream-fed be a result of our model somehow overestimating the
importance of the stream-fed mode?

It is not inconceivable that our model may somehow underestimate the merger rate at high redshift, even though the assumptions that it makes are quite standard.
The problem is that the entire main sequence of SFGs is shifted with respect to the data at $z=2$.
Mergers can temporarily increase the SFR by accelerating the conversion of gas into stars but they cannot change the average SFR for a given galaxy mass because
a shorter star formation timescale  means that at the end less fuel for star formation is left.

Let us now consider the second possibility, i.e. that the importance of stream-fed star formation is overestimated.
Even though our model incorporates disc instabilities,
in this article we have effectively identified stream-fed star formation with disc star formation and merger-driven star formation with bulge star formation.
In reality, while disc star formation is always stream-fed, bulge star formation is partly merger-driven and partly stream-fed.
By equating stream-fed star formation with disc star formation and merger-driven star formation with bulge star formation, we should underestimate the former and overestimate the latter, not the opposite, but let us analyse this point in closer detail.

Our semianalytic model has been run using the disc stability criterion (\ref{discinst}) for a disc instability threshold parameter of $\eta=0.7$.
In other words, we assume that, within the disc half-mass radius, the steady state is that the mass in the disc is 
half the total mass.
\citet{dekel_sari_ceverino09} have explored a model for the instability of clumpy discs and 
they have been able to relate the disc-to-total mass ratio $\delta$ within the the disc half-mass radius
to $\sigma/v_{\rm rot}$, the ratio of the velocity dispersion $\sigma$ within the disc to the disc's rotation speed $v_{\rm rot}$
(also see \citealp{genel_etal12}).
For a Toomre disc instability parameter $Q$ (\citealp{toomre64}; \citealp{binney_tremaine08}, Chapter~6) of $Q\sim 1$ and  a flat rotation curve, 
their model gives $\delta\sim \sqrt{2}\cdot\sigma/v_{\rm rot}$.
So, $\eta=0.7$ is equivalent to $\sigma/v_{\rm rot}\simeq 0.35$.
In fact, spectroscopic observations of gas kinematics in massive discs at $z\sim2$ give values closer to
$\sigma/v_{\rm rot}\simeq 0.2$ \citep{erb_etal04,foersterschreiber_etal06,cresci_etal09}, 
which imply $\delta\sim 0.3$.
Based on this analyis,
the actual inflow due to disk instability in high-$z$ clumpy discs
is stronger than what GalICS assumes.

Still, our predicted bulge-to-total stellar mass
ratios for blue galaxies are systematically on the high side at all masses
(GalICS predicts a typical bulge-to-total mass ratio of $M_{\rm bulge}/M_{\rm gal}\sim 0.3$ for a blue galaxy with $M_r \sim -20.5$;
  \citealp{cattaneo_etal06}, Fig.~10).
Simply lowering the value of $\eta$ would bring predicted bulge-to-total stellar mass ratios to values that are incompatible with the prevalence
of late-type morphologies in the local galaxy population.
This is why hydrodynamic simulations have for a long time experienced major difficulty in forming spirals with acceptable bulge-to-total stellar mass ratios,
though they follow gas and stellar dynamics more self-consistently than semianalytic models do.
Recent progress  in forming spirals with acceptable bulge-to-total stellar mass ratios 
is linked to strong feedback that preferentially ejects gas from the central starburst
\citep{governato_etal10,guedes_etal11,piontek_steinmetz11,brook_etal12,mccarthy_etal12}.
Therefore, a more physical description of disc instability must be accompanied by a more physical description of stellar feedback.

Based on this discussion, GalICS likely underestimates both the inflow into the bulge due to disc instability {\it and} the outflow from
the bulge due to stellar feedback. 
So, part of the stream-fed star formation that we predict to occur in the disc may have occurred in the bulge, or vice versa (as it is more likely, at least for spiral galaxies, whose bulge-to-stellar mass ratio are overpredicted).
However, that does not change the relative importance of stream-fed and merger-driven star formation.

One may also worry that our SFRs have been boosted by a factor of $(1+z)^{0.6}$ to produce more star formation at high redshift (Eq.~2; also see Fig.~8 of \citealp{cattaneo_etal06}
for the impact of this assumption on the evolution of the cosmic SFR density). 
While this assumption increases the stream-fed SFR,
the factor $(1+z)^{0.6}$ has been assumed to multiply both the disc SFR and the bulge SFR.
Thus it does not change the relative importance of the two in terms of star formation efficiency.

The most serious concern is that our model may exaggerate gas
consumption in protogalaxies at $z\gsim 4$.  with the consequence that
when the first major mergers occur, there is little gas left to
trigger a starburst.  To address this point, we have analysed the gas
content of major mergers.

 Fig.~\ref{merger_gasfraction} is a remake of Fig.~\ref{fig1} where we
 only show the galaxies that have experienced a major merger in the
 last $300\,$Myr (the circled ones) and compare them to the merger-like
   objects observed in Tacconi et al., (2008).  We colour-code
the symbols according to their gas fraction.  In GalICS, mergers of
 quenched, red-sequence galaxies  are gas poor, but
 that is due to our quenching criterion, which is related to the
 critical halo mass $M_{\rm crit}$, and has nothing to with the star
 formation efficiency.  So, let us focus on the galaxies that are on
 the main sequence of star forming galaxies.
 
 In our model, at $z=2$ and $M_*\lsim 10^{10.5}\,M_\odot$, gas
 fractions range from $\sim 0.3$ to $\sim 0.6$ There is a declining
 trend with mass, which is also observed (Tacconi et al. 2012).  At
 $M_*\sim 10^{11}\,M_\odot$, the typical gas fraction is $\sim
 0.15-0.2$, in agreement with the gas fractions of major mergers
 observed at $z=2-3$ \citep{tacconi_etal08}.  Therefore, there is no
 reason to believe that we underestimate the importance of
 merger-driven star formation because the gas fractions of our mergers
 are systematically lower than they should be. 

Our results are consistent with those that \citet{cattaneo_etal11} obtained with a much simpler galaxy formation model.
In that article we concentrated on mass assembly (here we concentrate on star formation).
We found that gas-rich mergers make a negligible contribution to the baryonic mass assembly of the overall galaxy population,
although they contribute about half of the mass in ellipticals with $M_*<10^{10.8}h^{-1}\,M_\odot$,
which formed their stars at lower $z$ than giant ellipticals due to downsizing \citep{thomas_etal05}. 
Elliptical galaxies with $M_*>10^{11}h^{-1}\,M_\odot$ accrete most of their mass
via dry mergers.
In Cattaneo et al. (2011), we did not consider the possibility that the critical mass above which gas accretion is shut down may be higher at higher redshifts and we wondered
to what extent that could affect our conclusions. Here we have included a redshift dependence of $M_{\rm crit}$ (Eq.~1), but the basic picture has not changed.

Lowering $z_{\rm crit}$, which is equivalent to increasing $M_{\rm crit}$ at high redshift,
 delays the shutdown of gas accretion but also transforms a number of dry mergers into wet mergers. 
The total amount of star formation increases but the relative importance of the two modes is almost unchanged.
Values of $z_{\rm crit}$ substantially lower than $z\sim 3$ cause GalICS to overpredict the galaxy luminosity function at $z=0$.
This effect can be partially compensated by lowering the value of $M_{\rm crit}$ at $z=0$ but there is limited leeway to do so.

In conclusion, the parameters of our model contain great
uncertainties (particularly those anchored to the luminosity function
of Lyman-break galaxies), which affect our capacity to trust the model's
predictions in quantitative detail, especially since we know that it
fails to reproduce the SFR-$M_*$ relation inferred from
observations at $z\sim 1-4$.

However, the basic results of our model are quite robust to changes in
parameter values and model assumptions because it is difficult to
increase the importance of wet mergers without also increasing the
importance of stream-fed star formation, unless one finds a way to
accrete large masses of gas while preventing them from making stars
until the first mergers occur.  These results corroborate the
scenario portrayed by \citet{dekel_etal09}, where they made a
distinction between the overall SFG population and extreme SMGs, which
are about ten times less numerous.  They conclude that the accretion
of cold gas is the main mode of galaxy formation in the former
population, whereas mergers are only necessary to explain the latter.

\bibliographystyle{mn2e}

\bibliography{ref_av}

\label{lastpage}
\end{document}